\documentclass[acmsmall,screen,review=FALSE, anonymous=FALSE]{acmart}

\usepackage{graphicx}
\usepackage{tcolorbox}
\usepackage{tikz}
\usepackage{tikz-qtree}
\usetikzlibrary{trees}
\usepackage{multirow}
\usepackage{multicol}
\usepackage{array}

\AtBeginDocument{%
  \providecommand\BibTeX{{%
    \normalfont B\kern-0.5em{\scshape i\kern-0.25em b}\kern-0.8em\TeX}}}

\setcopyright{acmcopyright}
\acmJournal{PACMHCI}
\acmYear{2022}
\acmVolume{6}
\acmNumber{CSCW2}
\acmDOI{10.1145/nnnnnnn}
\usepackage{csquotes}

\begin{document}

\title{The Data-Production \textit{Dispositif}}

\author{Milagros Miceli}
\email{m.miceli@tu-berlin.de}
\orcid{0000-0003-0585-3072}
\affiliation{%
  \institution{DAIR Institute, TU Berlin, and Weizenbaum Institute}
  \country{Germany \& USA}
}

\author{Julian Posada}
\email{julian.posada@mail.utoronto.ca}
\orcid{0000-0002-3285-6503}
\affiliation{%
  \institution{University of Toronto \& Schwartz Reisman Institute}
  \country{Canada}
}

\begin{abstract}
Machine learning (ML) depends on data to train and verify models. 
Very often, organizations outsource processes related to data work (i.e., generating and annotating data and evaluating outputs) through business process outsourcing (BPO) companies and crowdsourcing platforms. 
This paper investigates outsourced ML data work in Latin America by studying three platforms in Venezuela and a BPO in Argentina. We lean on the Foucauldian notion of \textit{dispositif} to define the \textit{data-production dispositif} as an ensemble of discourses, actions, and objects strategically disposed to (re)produce power/knowledge relations in data and labor. Our dispositif analysis comprises the examination of 210 data work instruction documents, 55 interviews with data workers, managers, and requesters, and participant observation. Our findings show that discourses encoded in instructions reproduce and normalize the worldviews of requesters. Precarious working conditions and economic dependency alienate workers, making them obedient to instructions. Furthermore, discourses and social contexts materialize in artifacts, such as interfaces and performance metrics, limiting workers’ agency and normalizing specific ways of interpreting data. We conclude by stressing the importance of counteracting the data-production dispositif by fighting alienation and precarization, and empowering data workers to become assets in the quest for high-quality data.
\end{abstract}



\maketitle

\section{Introduction}
Many machine learning (ML) models are built from training data previously collected, cleaned, and annotated by human workers. Companies and research institutions outsource several of these tasks through online labor platforms \cite{Posada2020a} and business process outsourcing (BPO) companies \cite{miceli2020}. In these instances, outsourcing organization and their clients regard workers as independent contractors, considering them factors of production, and their labor a commodity or a product subject to market regulations \cite{Wood2019a}. They are paid as little as a few cents of a dollar per task, usually lack social protection traditionally tied with employment relations, and are subject to systems of control and surveillance \cite{Tubaro2020,Gray2019, Irani2015}. Their assignments broadly comprise the interpretation and classification of data, and their work practices involve subjective social and technical choices that influence data production and have ethical and political implications.  Workers interpreting and classifying data do not do so in a vacuum: their labor is embedded in large industrial structures and deeply intertwined with naturalized profit-oriented interests \cite{kazimzade_biased_2020}. 

This paper presents an investigation of data production for ML as carried out by Latin American data workers mediated by three platforms operating in Venezuela and a business process outsourcing (BPO) company located in Argentina. To study \textit{data work} for machine learning, which we define as the labor involved in the collection, curation, classification, labeling, and verification of data, we lean on Foucault's notion of \textit{dispositif} and apply the method of \textit{dispositif analysis} \cite{jager2016analysing}. A dispositif is an ensemble of objects, subjects, discourses, and practices as well as the relations that can be established between them \cite{foucault1980}. Examples of dispositifs include prisons, police, and academia. These heterogeneous ensembles of discursive and non-discursive elements constitute what is perceived as reality and, as such, what is taken for granted. 

\begin{figure}[h]
    \centering
    \includegraphics[scale=0.25]{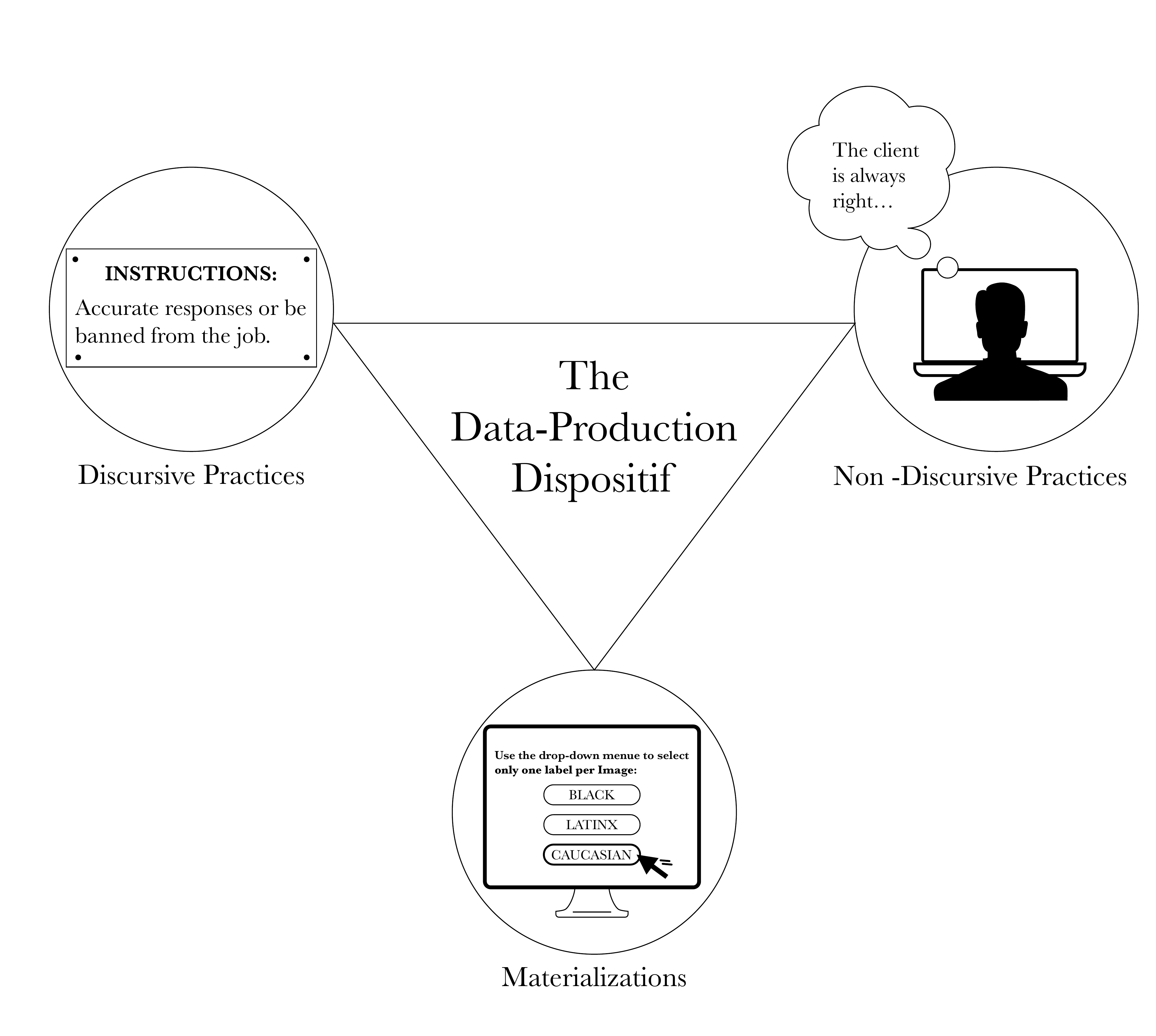}
    \caption{The three components of the data-production dispositif based on the framework and figure proposed by Jäger and Maier \cite{jager2016analysing}}
    \label{fig:dispositif}
\end{figure}

The decision to lean on Foucault's notion of dispositif is methodological, rather than theoretical. This notion and the method of dispositif analysis enables the study of data production as embedded in social interactions and hierarchies that condition how data is constructed and how specific discourses are reproduced. As we will describe in Section 3.1, this method also allowed us to integrate diverse qualitative data and focus on the relationships between them. 
We define the \textit{data-production dispositif} as the network of discourses, work practices, hierarchies, subjects, and artifacts comprised in ML data work (see Figure \ref{fig:dispositif}) and the power/knowledge relationships that are established and naturalized among them. The data-production dispositif determines the realities that ML datasets can reflect and the ones that remain erased from them. It has a crucial effect on the outputs that ML models will consider to be true. Dispositif analysis interrogates means of reality making, with a special focus set on the meanings that become dominant and those that are marginalized\,---\,“the said as much as the unsaid” \cite{foucault1980}. Our dispositif analysis explores the sites where the production of ML data is outsourced. It comprises the investigation of (1) \textit{linguistically-performed elements} (what is said/written), (2) \textit{non-linguistically performed practices} (what is done), and (3) \textit{materializations} (how linguistically and non-linguistically performed practices translate into objects) \cite{jager2016analysing}.  

These elements and research questions relate specifically to the outsourcing of ML data-production tasks and can be structured as follows:
\begin{itemize}
    \item 
    \textbf{Linguistically performed elements}: \emph{What discourses are present in task instructions provided to outsourced data workers?} (RQ1)
    
     $\rightarrow$ We analyzed a corpus of 210 instruction texts for data-related tasks requested by ML practitioners and outsourced to data workers. 
    \item \textbf{Non-linguistically performed practices}: \emph{How do outsourced data workers, managers, and requesters interact with each other and instruction documents to produce data?} (RQ2)
    
     $\rightarrow$ To explore how linguistically performed elements translate into practice, we conducted 41 interviews with data workers and inquired how they interpret the instructed tasks. In addition, we conducted interviews with six managers and eight ML practitioners (in their role as data-work requesters).   
    \item \textbf{Materializations}: \emph{What artifacts support the observance of instructions, and what kind of work they perform?} (RQ3)
    
     $\rightarrow$ Through participant observation, we account for some of the material elements in which the data-production dispositif manifests, such as platforms and interfaces, tools to surveil workers, and documents that record the decisions made between service providers and service requesters.
\end{itemize}

To summarize, this paper, its contribution, and the extensive analysis it comprises can be described as follows:
We start by exploring Foucault’s notion of dispositif and defining key related concepts such as power, knowledge, and discourse. 
Then, we review previous investigations that have discussed ML data work and further define the scope of the data-production dispositif. 
After offering an overview of the dispositif analysis method, informants, and fieldwork sites, we present our findings. 
They are organized around the three elements that form the dispositif. 

The findings show that, instead of seeking the “wisdom of crowds,” where a diverse and independent group cooperates to solve a problem, requesters use task instructions to impose predefined forms of interpreting, classifying, and sorting data that respond primarily to profit-oriented interest. Managers in BPOs and algorithms in labor platforms are in charge of overseeing the process. Poverty and dependence in the areas where data work is outsourced leaves workers with no other option but to obey and avoid questioning instructions. Documents, tools, and interfaces constitute some of the dispositif's materializations. 
Given these findings, we outline some implications and propose three ways of counteracting the current composition of the data-production dispositif and its effects by fighting workers’ precarization, alienation, and surveillance. Finally, we discuss the limitations of our investigation.

\section{Defining Key Concepts}

\subsection{\textit{Dispositif} and Other Foucauldian Concepts}

Foucault proposes a relational conception of \emph{power} \cite{foucault1980} and argues that its exercise takes place in networks of relations rather than being placed in a specific social location \cite{cronin1996}. He defines power as “a whole series of particular mechanisms, definable and defined, that seem capable of inducing behaviours or discourses”\cite{Foucault1996-FOUWIC}.  Therefore, power is not held by or exercised over individuals but works through the impersonal relations of force and strategy that connect subjects \cite{cronin1996}.  
Power operates through practices that act upon subject's present or future actions. It is effective as long as it is normalized, that is, taken for granted and perceived as the inevitable way things are. 

Present throughout Foucault's power analysis is the implicit relationship between power and knowledge, whereby one implies the other. He understands \emph{knowledge} as entangled with discursive power and describes it as the power to define others and to produce truth through discourses. Knowledge and power are integrated with one another because “it is not possible for power to be exercised without knowledge, it is impossible for knowledge not to engender power” \cite{foucault1980}. 

In a related manner, Foucault uses the term \emph{discourse} to refer to a historically contingent system that produces knowledge and meaning \cite{foucault1982a}. Discourse is not only a way of organizing and presenting knowledge, but also, it can structure social practices and the relations that emerge through the collective understanding of reality encoded in discourse \cite{foucault1971}. 
Discourses encode power in the sense that they can determine reality. Subjects are active participants in reality-making processes as co-producers of discourses, which puts explicit and implicit knowledge at their disposal \cite{jager2016analysing}. Thus, discursive power is not exercised \emph{on} subjects but flows \emph{through} them. 
However, discourses still have a disciplinary effect on subjects as they assure the prevalence of certain knowledge of what can be said, done, and thought.

Power, knowledge, and discourse finally converge in \emph{dispositif}, a notion that expands discourse to include non-discursive practices and artifacts. Foucault defines dispositif as “a thoroughly heterogeneous ensemble consisting of discourses, institutions, architectural forms, regulatory decisions, laws, administrative measures, scientific statements, philosophical, moral and philanthropic propositions – in short, the said as much as the unsaid … The [dispositif] itself is the system of relations that can be established between these elements” \cite{foucault1980}. There is a multiplicity of dispositifs that influence each other and have strategic functions within power relationships. As Foucault puts it, dispositifs respond to an “urgent need” that is bound to specific historical and geographical contexts. 
Link \cite{link2014} draws attention to the etymological root shared by the French words \textit{“disposition”} and \textit{“dispositif.”} In colloquial French, “disposition” is used in the sense of being “at someone's \emph{disposal}.” This way, Link highlights the power element comprised in dispositif as the separation between those who are “at the disposal” (those who are instrumental) and those who have influence to determine the strategy used to meet a need \cite{caborn2016}. 
Those who are “at the disposal” and those who “dispose” are part of the dispositif's strategy. 

In sum, the concept of dispositif comprises the knowledge that is built into linguistically performed practices (what is said, written, and thought), non-linguistically performed practices (what is done), and materializations (the objects) \cite{jaeger2007, jager2016analysing}. A dispositif can therefore be defined as a constantly changing network of objects, subjects, discourses, and practices that shape each other, producing new knowledge and new power. 

Previous HCI and CSCW research has engaged with Foucauldian theory.  For instance, Harmon and Mazmanian \cite{harmon2013} follow Foucault's understanding of \textit{discourse} to explore how US residents talk about smartphones and smartphone users.  Kou et al. \cite{kou2019} use the Foucauldian concepts of \textit{power}, \textit{knowledge}, and \textit{self} to explicate human-technology relationships. And Bardzell et al. \cite{bardzell2014} draw on Foucault's Theory of Identity to investigate social practices within the virtual world Second Life. In terms of methodology,  Kannabiran et al. \cite{kannabiran2011} use Foucauldian Discourse Analysis (FDA) to study the rules and mechanisms involved in HCI discourses on sexuality, while Spiel \cite{spiel2017} combines Actor-Network Theory with FDA into a “critical experience” framework to evaluate how children in the autistic spectrum interact with technologies.  Despite the wide application of Foucauldian theory in HCI and CSCW, the dispositif notion and analysis has not been applied to the study of data production and data work. To that end, we believe that our contribution could be seen as methodological, in the sense that we intend to show a novel and comprehensive mode of analysis to approach data work.

\subsection{Data Work for Machine Learning}

The data-production dispositif analyzed in this paper comprises the “infrastructure” that enables the (re-)production and circulation of specific discourses in and through ML data work.  As Foucault argues, the emergence of each dispositif responds to an “urgent need.”  The data-production dispositif responds to the growing demand for data and labor in the AI industry. 

We define \textit{data work} as the human labor necessary for data production, in this case, for machine learning. Data work involves the collection, curation, classification, labeling, and verification of data. Users, developers, and outsourced workers carry out these tasks at any point in the development and deployment of AI systems. For example, medical professionals in the case of AI for healthcare \cite{Thakkar2022, Moller2020, bossen2019}, education professionals \cite{Lu2021}, or internet users when answering ReCAPTCHA tests \cite{Justie2021}.  
This paper will employ the term “data work” to refer exclusively to the labor outsourced through crowdsourcing platforms and specialized business process outsourcing (BPO) companies, instead of the broader data work carried out by other professionals and users, while acknowledging the role of the former within the dispositif as requesters. 

Platforms are one of the two significant ways of outsourcing data work. The rise of alternative forms of work different from traditional employment \cite{Katz2016} and the expansion of the “gig economy,” or casual employment mediated through platforms \cite{Woodcock2020}, gave rise to “crowdsourcing,” “crowdwork,” or “digital piecework” platforms where geographically dispersed workers are allocated many fragmented tasks, which are carried out online from their homes. Platforms are hybrid organizations that combine traits of firms and multi-sided markets \cite{Casilli2019}. They serve as infrastructures that “facilitate and shape personalised interactions among end-users and complementors, organised through the systematic collection, algorithmic processing, monetisation, and circulation of data” \cite{Poell2019}. Platforms thrive in digital environments because they respond to deficiencies in markets and enterprises that fail to extract and appropriate data and allocate resources efficiently \cite{Casilli2019}. 

The second primary form of outsourced data work for ML is provided by business process outsourcing (BPO) companies.  Conversely to crowdsourcing platforms, where hierarchies are primarily managed by algorithms, BPOs show rather traditional management structures. BPO is a form of outsourcing that involves contracting a third-party service provider to carry out specific parts of a company’s operations, in the case of our investigation, data-related tasks. These service providers often specialize in one type of ML data service (e.g., semantic segmentation) or application domain (e.g., computer vision), contrary to platforms specializing in one or a few application domains but with more diverse data services. While prices per piece are significantly higher than those offered by platforms, many machine learning companies prefer to outsource their data-related projects with BPOs because of the perceived higher quality of data \cite{miceli2021a}. This is due to the companies’ domain specialization and traditional managerial structures that allow more direct and personal communication. 

The intervention of humans in processes of data production has been addressed by a large body of CSCW and HCI research \cite{passi2017, passi2018b, muller2019, muller2021, miceli2020, feinberg2017, seidelin2018, gadiraju2017,Thakkar2020}. Some investigations have explored the role of worker subjectivity on datasets \cite{brodley1999, cheng2013, wauthier2011, ghai2020} and have proposed ways to recognize and address worker bias \cite{geva2019, artstein2005, hube2019, wauthier2011}. In contrast, other researchers have documented the “practices, politics, and values of workers in the data pipeline” \cite{Sambasivan2022,Sambasivan2021} and the sociotechnical organization of data work that privileges speed, scale, and scalability over worker wellbeing \cite{Irani2015}, low wages \cite{hara2018, fan2020}, dependency \cite{ross2010}, and the power asymmetries vis-à-vis requesters \cite{irani2013, salehi2015a, martin2014, miceli2020}.

As we argue, the “urgent need” addressed by the data-production dispositif is the exponential need for cheaper and more profitable data, which is also the exploitation of surveillance \cite{Zuboff2019}, natural resources \cite{ Dauvergne2020}, and other types of labor \cite{Crawford2021}. Previous research has highlighted the role of these elements to guarantee a façade where AI is seen as neutral, unbiased, and efficient due to the lack of human intervention \,---\,and error\,---\,while keeping workers and factors of production hidden from the public lens \cite{Casilli2017a, Irani2015, Gray2019}. These elements show the wide extension of the “heterogeneous ensemble” that constitutes the discursive, non-discursive, and material elements of data production. Because the data-production dispositif is too vast to explore in one academic paper, we circumscribe its analysis around outsourced data work for ML as one of its crucial components.

\section{Methodology}

\subsection{Dispositif Analysis of Data Production}

We lean on dispositif analysis to investigate the discourses implicit in annotation instructions, the non-discursive practices involved in the production of ML datasets, and how both materialize in artifacts. Often described as an extension of discourse analysis \cite{caborn2016}, dispositif analysis expands the field of inquiry beyond texts to include actions, relationships, and objects. Dispositif analysis rests on the notion of knowledge (and power) as the connecting force between discursive and non-discursive components. It accounts for hierarchies and power structures in societal fields and organizations that shape the construction of meaning in discourse \cite{raffnsoe2016}. Thus, our dispositif analysis crucially focuses on the relationship between discourse, practice, and objects in data production, and the power created through their interaction.

Foucault never outlined an explicit methodology of dispositif analysis. Several authors, most prominently Sigfried Jäger \cite{jaeger2007,jager2016analysing}, have explored ways of operationalizing the complex Foucauldian notion of dispositif into a method of inquiry. Dispositif analysis has thus been in constant evolution since the mid-1980s. Caborn \cite{caborn2016} mentions four steps comprised in this methodology: (1) identifying the elements that constitute the dispositif, (2) determining which discourses they embody and their entanglement with other discourses, (3) interrogating power by “considering who or what is \textit{at the disposal} of whom”, and (4) analyzing non-discursive practices associated with the dispositif's discourses. However, to our knowledge, a comprehensive guide or method of \emph{how to} conduct a dispositif analysis has not yet been developed. As Jäger and Maier describe, dispositif analysis remains “a flexible approach and systematic incitement for researchers to develop their analytic strategies, depending on the research question and type of materials at hand” \cite{jager2016analysing}. 

The study presented in this paper follows the experimental spirit of Jäger and Maier's invitation to develop our analytical strategy, in this case, to study data production. Here, we combine methodological elements discussed by several authors in terms of the operationalization of power, knowledge, and discourse \cite{buehrmann2007,caborn2016,larroche2019, link2014, nowicka-franczak2021}, apply them to our fieldwork on data production through platforms and at a BPO, and follow the examples provided by previous research that has successfully applied variations of dispositif analysis \cite{caborn2016, hamann2019, manderscheid2014, whelan2019, wichum2013}. We followed the four steps outlined by Caborn mentioned above and based our analysis on the three-dimensional framework described by Jäger and Maier \cite{jager2016analysing} as follows:

\begin{itemize}
    \item The \emph{analysis of linguistically performed elements}: which aims at reconstructing the knowledge built into what is said and written through discourse analysis. In terms of our investigation, this phase comprised an examination of the discourses encoded in the instruction documents received by data workers. 
    \item The \emph{analysis of non-linguistically performed practices}: which aims at reconstructing the knowledge that underlies linguistically performed practices and how they translate into action. In this phase, we investigated how workers make sense of the task instructions and their work in general, the interactions between workers, managers, and clients, and the labor conditions that structure these practices. We studied these elements through interviews conducted with data annotators who perform tasks guided by such instructions, machine learning practitioners who compose annotation instructions, and managers who oversee the process.
    \item The \emph{materializations}: This phase of analysis consisted of identifying the knowledge that is built into physical and digital artifacts, i.e., discursive materialization, whose existence is coherent with the discourses they encode. Through this lens, we set the focus on the platforms and interfaces used to perform data work, documents (as artifacts and not as texts) that record decision-making processes, and tools used to surveil workers and quantify their performance. Our analysis of these materalizations is based on participant observations and the above-mentioned interviews.
\end{itemize} 

\subsection{Researcher Positionality}

Making researchers' positionality explicit is key to situate the standpoint from which an investigation has been conducted. Positionality statements are relevant to all types of studies, specially qualitative and exploratory investigations such as this one. Moreover, given the flexible character of dispositif analyses, it seems appropriate to disclose some elements of the authors' backgrounds that might have informed the analysis presented in this paper. 

Both authors are multiracial researchers born in different countries of Latin America. Both are first-generation academics working in institutions located in the Global North, where they live under immigrant status. Both have a background in Sociology and Communication. Their first language is Spanish. One of the authors identifies as female and the other as male. Both are cisgender. Despite being born and raised within worker-class families and in the same regions as the data workers interviewed, the authors acknowledge that their class-related experiences differ from those of the interview partners and that their position as researchers living and working in the Global North provides the authors with privilege that the study participants do not hold. Throughout data collection, analysis, and while considering the implications of this investigation, the authors have put much effort in remaining reflexive and acknowledging their position regarding the study participants and field of inquiry.

\subsection{Data Collection and Analysis}

This investigation comprises several weeks of participant observation, a total of 55 interviews, and the analysis of 210 instruction documents. These data were collected during several months of fieldwork from 2019 to 2021, online and in person, at two sites (see Table \ref{tab:sites}): 
\begin{itemize}
    \item virtually studying three crowdsourcing platforms operating in Venezuela and the experiences of platform workers, and,
    \item in a hybrid format, at a business process outsourcing company (BPO) located in Buenos Aires, Argentina, where data workers perform tasks related to the collection and labeling of data for machine learning.
\end{itemize}
At both fieldwork sites, we conducted participant observations and semi-structured interviews. To complement these data, we conducted a series of expert interviews with managers at other BPOs and with ML practitioners in their role of data-work requesters (see Section 3.3.2 and Table \ref{tab:interviews} for a detailed account of the interview participants).

\subsubsection{Fieldwork}
\hspace{1cm}

Fieldwork in Venezuela was carried out virtually between July 2020 and June 2021 due to restrictions related to the coronavirus pandemic. For the first phase of this research, we signed up and completed tasks for the platforms to understand the tasks available, working conditions, and interfaces. While some of these platforms presented similar tasks, they differed considerably in their general availability, the interfaces, labor process, and task applications. Initially, we contacted the platform workers using convenience sampling since this population is invisible, meaning that it is difficult to approach them without the support of the platforms, which we did not have for the study. We sought permission from the moderators of the most popular worker groups on Facebook and Discord to post a call for study participants. Thanks to this initial approach, we were able to use snowball sampling to contact further participants. We conducted in-depth interviews and asked workers about their experience working for the platforms. Additionally, we asked workers if they could share information about the instructions they received. Some workers also shared guides created by colleagues to understand and answer the tasks efficiently. We include those in our analysis as well. Moreover, we also searched the internet to find additional annotation instructions online. Our approach, notably the use of convenience and snowball sampling, present several limitations in terms of reproducibility and bias towards participants belonging to similar social circles. We have mitigated these issues by comparing our results with similar studies (see Section 2.2) and with the workers at the BPO company.

Fieldwork at Alamo, the Argentine business process outsourcing (BPO) company, was carried out in-person between May and June 2019 in Buenos Aires and continued online between August 2020 and February 2021. At the time of this investigation, this company is a medium-sized organization with branches in several Latin American countries. 
Besides data work, Alamo conducts content moderation and software testing projects.  The company is an impact sourcing type of BPO, which refers to a branch of the outsourcing industry that purposely employs workers from poor and marginalized populations to offer them a chance in the labor market and to provide information-based services at lower prices.
We contacted the company via e-mail to request field access. After several months of inquiry, a meeting with the company's management took place in which the researcher on site signed a non-disclosure agreement that specified several elements that we are not allowed to disclose in this or other papers. Most of these elements concern the identity of clients and specific details about their ML models. After this meeting, fieldwork was allowed to commence and we were able to observe several projects related to the collection and annotation of data for ML. Apart from shadowing workers, we were granted access to team meetings, meetings with clients, workers' briefings, and QA analysis related to three projects carried out by the company in 2019, involving the collection and labeling of image data. We complemented the observations with in-depth interviews with data workers, managers, and QA analysts. 

\begin{table}
\caption{Fieldwork sites: Studied data work BPO and platforms}
\label{tab:sites}
\tiny
\def\arraystretch{1.2}
\resizebox{\textwidth}{!}{%
\begin{tabular}{p{2cm}p{1.5cm}p{4cm}p{3cm}}
\cline{1-4} \cline{1-4}
\textbf{Entity} & \textbf{Type} & \textbf{Primary Tasks} & \textbf{Applications} \\ 
\hline
\textbf{ALAMO} & BPO & \begin{tabular}[c]{@{}l@{}}Data collection and annotation\\ Content moderation\end{tabular} & E-commerce \\ 
\hline
\textbf{CLICKRATING} & Platform & \begin{tabular}[c]{@{}l@{}}Data entry\\ Algorithmic verification\end{tabular} & Online search engine \\ \hline
\textbf{TASKSOURCE} & Platform & \begin{tabular}[c]{@{}l@{}}2D/3D image classification\\ 2D/3D semantic segmentation\end{tabular}
& \begin{tabular}[c]{@{}l@{}}Self-driving vehicles\\ Internet of things\end{tabular} \\
\hline
\textbf{WORKERHUB} & Platform & \begin{tabular}[c]{@{}l@{}}2D image, text and video classification\\ 2D semantic segmentation\\ Text transcription\end{tabular} & \begin{tabular}[c]{@{}l@{}}Content moderation\\ E-commerce\\ \end{tabular} \\ 
\cline{1-4}
\cline{1-4}
\end{tabular}%
}
\end{table}

\hspace{1cm}

\subsubsection{Instruction Documents}
\hspace{1cm}

\begin{table}
\caption{Evolution of codes throughout the three phases of discourse analysis as applied to the instruction documents}
\label{tab:codes}
\tiny
\def\arraystretch{1.7}
\resizebox{\textwidth}{!}{%
\begin{tabular}{p{1cm}p{1.9cm}p{1.9cm}p{6cm}}
\cline{1-4} \cline{1-4}
\multicolumn{1}{c}{\textbf{Structural Analysis}} &
  \multicolumn{1}{c}{\textbf{Detailed Analysis}} &
  \multicolumn{1}{c}{\textbf{Synoptic Analysis}} &
  \multicolumn{1}{c}{\textbf{Memo}} \\ \hline
\multirow{5}{1cm}{Document characteristics} &
  Format &
  \multirow{3}{1.9cm}{Documents as constraints} &
  \multirow{3}{6cm}{Document elements that hinder the correct understanding or completion of tasks. Elements that constrain workers’ interpretation of data.} \\ \cline{2-2}
 &
  Document language &
   &
   \\ \cline{2-2}
 &
  Language barriers &
   &
   \\ \cline{2-4}
 &
  Document version &
  \multirow{2}{1.9cm}{Document as artifacts} &
  \multirow{2}{6cm}{Evidence of documents evolving / being modified by requesters. Different versions of same task.} \\ \cline{2-2}
 &
  1st Person &
   &
   \\ \cline{1-4}
\multirow{3}{1cm}{Descriptions} &
  Project description &
  \multirow{3}{1.9cm}{Worker alienation} &
  \multirow{3}{6cm}{Evidence of workers kept in the dark about the ML pipeline, made to feel foreign to the products of their labor. Workers forced to automize their outputs.} \\ \cline{2-2}
 &
  AI description &
   &
   \\ \cline{2-2}
 &
  Use of tech jargon &
   &
   \\ \cline{1-4}
\multirow{11}{1cm}{Task type} &
  Data collection &
  \multirow{3}{1.9cm}{Data generation} &
  \multirow{3}{6cm}{Data scraping, collection. Tasks that include taking or intervening pictures and/or generating texts.} \\ \cline{2-2}
 &
  Classification &
   &
   \\ \cline{2-2}
 &
  Involves text &
   &
   \\ \cline{2-4}
 &
  Data labeling &
  \multirow{3}{1.9cm}{Data annotation} &
  \multirow{3}{6cm}{Data segmentation and classification.  Task related to labeling and keywording.} \\ \cline{2-2}
 &
  Segmentation &
   &
   \\ \cline{2-2}
 &
  Involves images &
   &
   \\ \cline{2-4}
 &
  Keywording &
  \multirow{3}{1.9cm}{Algorithmic verification} &
  \multirow{3}{6cm}{Assesment of algorithmic output by workers. Testing of ML systems, rating of search engine queries, moderation of content flagged by algorithm.} \\ \cline{2-2}
 &
  Evaluation &
   &
   \\ \cline{2-2}
 &
  Testing &
   &
   \\ \cline{2-4}
 &
  Rating &
  \multirow{2}{1.9cm}{Algorithmic impersonation} &
  \multirow{2}{6cm}{Workers are instructed to act as an AI and rendered invisible in the process.} \\ \cline{2-2}
 &
  Content moderation &
   &
   \\ \cline{1-4}
\multirow{4}{1cm}{Taxonomies} &
  Categories &
  Multiclass classification &
  Classifications that include more than two options. Rationale behind classes. Normalization. \\ \cline{2-4}
 &
  Classes &
  Binary classification &
  Binary classifications and simplification of complex phenomena. \\ \cline{2-4}
 &
  Atributes &
  Other/Unknown &
  Use of a third label, especially in relation to binary classifications. Otherness. \\ \cline{2-4}
 &
  Labels definition &
  Ambiguity/Exceptions &
  How ambiguity and marginal cases are dealt with. \\ \cline{1-4}
\multirow{5}{1cm}{Examples} &
  Example description &
  Errors/Discrepancies &
  Cases where instructions contain errors.  Discrepancies between instruction and interface. \\ \cline{2-4}
 &
  Counter-example &
  \multirow{2}{1.9cm}{Explicit} &
  \multirow{2}{6cm}{What is explicitly described.  Rationale behind taxonomies and examples made explicit.} \\ \cline{2-2}
 &
  Clarifications &
   &
   \\ \cline{2-4}
 &
  Interface description &
  \multirow{2}{1.9cm}{Implicit} &
  \multirow{2}{6cm}{What remains unsaid. What is considered self-evident. Implicit rationale behind taxonomies.} \\ \cline{2-2}
 &
  Use of images &
   &
   \\ \cline{1-4}
\multirow{2}{1cm}{Disturbing content} &
  Content warnings &
  \multirow{2}{1.9cm}{Exposure to disturbing content} &
  \multirow{2}{6cm}{Tasks that include dealing with that is sexual or violent in nature.} \\ \cline{2-2}
 &
  Content inclusion &
   &
   \\ \cline{1-4}
\multirow{4}{1cm}{Worker skills} &
  Experience &
  \multirow{2}{1.9cm}{Language} &
  \multirow{2}{6cm}{Discrepancies between instructions language and workers' native language. Problems and strategies. Use of Google Translate. Guides developed by workers in Spanish.} \\ \cline{2-2}
 &
  Language skills &
   &
   \\ \cline{2-4}
 &
  Technical skills &
  \multirow{2}{1.9cm}{Quantification and surveilance} &
  \multirow{2}{6cm}{References to how the performance of workers is measured and surveiled. Consequences for low scores.} \\ \cline{2-2}
 &
  Score &
   &
   \\ \cline{1-4}
\multirow{4}{1cm}{Warnings} &
  Ban &
  \multirow{2}{1.9cm}{Worker obedience} &
  \multirow{2}{6cm}{How the unquestioning obedience is fostered in task instructions. How workers are prompted to think in terms of what the requester wants.} \\ \cline{2-2} 
 &
  Speed &
   &
   \\ \cline{2-4}
 &
  No payment &
  \multirow{2}{1.9cm}{Worker precarization} &
  \multirow{2}{6cm}{Pracarious labor conditions made explicit in instruction documents. Threats of being banned from the task. References to piece rate pay. Arbitrary deffinitions of worker accuracy.} \\ \cline{2-2}
 &
  Accuracy &
   & 
    \\ \cline{1-4} \cline{1-4}
\end{tabular}%
}
\end{table}

In total, we collected 210 annotation instruction documents from the platforms and the BPO. The analysis of the instructions was carried out by both authors. We used critical discourse analysis \cite{jager2016analysing} to explore the instruction texts. The analysis comprised three stages: (1) the structural analysis of the corpus, (2) a detailed analysis of discourse fragments, and (3) a synoptic analysis. These steps (especially the synoptic analysis) included several iterations that allowed us to discover connections between different levels of analysis, collect evidence to support our interpretations, and develop arguments. Table \ref{tab:codes} offers an overview of the codes used for the discourse analysis of the instruction documents, their evolution throughout the three phases of analysis, and explanatory memos that reflect our understanding of each code.

The goal of the \emph{structural analysis} is to code the material to identify significant patterns and recurring themes and sub-themes comprised in the instructions.  By the end of the structural analysis phase, we were able to identify elements of the text structure, regular tasks, and stylistic devices that appeared in the instruction documents. These elements helped us identify “typical” texts and representative discourse fragments for the following analysis step.

The \emph{detailed analysis} comprised an examination of selected text fragments.  We focused on identifying typical representations and their variations and interrogated the elements highlighted in the instruction documents and the contextual knowledge that is taken for granted and, thus, neglected in them. We also paid special attention to binary reductionisms, presupposition and attribution, examples, and visualizations. A critical aspect of the analysis focused on the taxonomies that structure the labels instructed by requesters. 
 
Finally, the \emph{synoptic analysis} included the overall interrogation of the observations that emerged from the structural and detailed analyses. This phase included an intensive exchange between both authors to reflect upon our shared understanding of the identified discourse strands. We contrasted the selected fragments and the identified elements with the interview and observation material. This approach helped us understand the role of work and managerial practices in legitimizing specific discourses.

\subsubsection{Interviews}
\hspace{1cm}

To reconstruct the knowledge that underlies the practices that constitute the data-production dispositif, we turned to the experiences of those actors who interact with the instructions regularly. With this aim, we conducted a total of 55 interviews with dataworkers located in Venezuela and Argentina, BPOs managers and founders, and ML practitioners who regularly outsource data-related tasks. Table \ref{tab:interviews} shows a detailed overview of the interview partners, including their role within their organizations, location, type of interview, and language.

Due to the restrictions related to the COVID-19 pandemic, the interviews with platform workers were conducted online through video calls, while those with BPO workers were conducted in person before the pandemic. The interviews with the data workers were conducted in Spanish, which is the native language of the interviewers and the participants. We conducted the expert interviews in English. While most platform workers had tried several platforms, they usually focused on one, except for two workers who worked simultaneously for Tasksource and Workerhub. All interview partners were asked to choose a code name or were anonymized post-hoc to preserve their identity and that of related informants.

The goal of the interviews was to reveal practices and perceptions and obtain additional information about the organizational relations and structures that inform how data-related tasks come about, how instructions are communicated, and how workers execute them. For instance, hierarchical structures can have an essential effect on meaning-making practices as enacted through the annotation instructions without being referred to explicitly or implicitly in the instruction documents. The in-depth interviews with data workers include accounts of specific work situations involving the interpretation of data. Moreover, they cover task descriptions, widespread routines, practices, working conditions, lived experiences, and general views on their work and the local labor market. 

It is essential to mention that the differentiation between in-depth and expert interviews refers to the interview method chosen for each situation and informant and was not based on informants' occupational status or position. Our priority was engaging in in-depth conversations with data workers to discuss and learn from their experiences in and beyond data work. Conversely, we used the expert-interview method to conduct focused exchanges with actors that possessed a broad overview of the machine learning pipeline. The expert interviews covered the topics of data work and the relationship between BPO/platform and requesters.  

Dispositif analysis allowed us enough flexibility to obtain valuable insights from the interviews by combining inductive and deductive coding. Some of the topics that we identified through discourse analysis in the instruction documents helped us build categories to code the interviews. This form of deductive coding was oriented towards finding additional evidence for phenomena identified in the instruction texts and understanding the contexts in which instructions are formulated and carried out. In addition, there was room for inductive category formation so that several codes could emerge directly from the interviews during coding. This approach helped us identify valuable observations that otherwise have been lost. 
Through this form of analysis, we aimed at identifying patterns. Those patterns were later confronted with the elements identified in the instruction texts and complemented with participant observations. 
The development of coding schemes for the analysis and the coding process itself was carried out in iterations involving cross-coding between both authors. The interview transcripts were analyzed in their original language (Spanish or English). The excerpts included in Section \ref{sec:4} were translated by us when writing this paper and only after the analysis phase.
Our emergent understanding evolved throughout numerous discussions and several iterations until reaching the set of findings that we present in Section \ref{sec:4}.

\begin{table}
\caption{Overview of interview partners and interview characteristics}
\label{tab:interviews}
\def\arraystretch{1.2}
\resizebox{\textwidth}{!}{%
\begin{tabular}{p{2.2cm}p{5.3cm}p{2.7cm}p{1.4cm}p{1.4cm}p{3.5cm}} \cline{1-6} \cline{1-6}
& \multicolumn{1}{c}{\textbf{ORGANIZATION}} & \multicolumn{1}{c}{\textbf{INTERVIEW METHOD}} & \multicolumn{1}{c}{\textbf{MEDIUM}} & \multicolumn{1}{c}{\textbf{LANG.}} & \multicolumn{1}{c}{\textbf{INFORMANTS}} \\ \hline
\multirow{4}{2cm}{\textbf{WORKERS}} & Alamo (BPO in Argentina)            & In-depth interview        & In-person       & Spanish           & 10 data workers                                                     \\ \cline{2-6}
                                                          & Tasksource (Platform in Venezuela)  & In-depth interview        & Zoom            & Spanish           & 8 data workers                                                      \\ \cline{2-6}
                                                          & Workerhub (Platform in Venezuela)   & In-depth interview        & Zoom            & Spanish           & 19 data workers                                                      \\ \cline{2-6}
                                                          & Clickrating (Platform in Venezuela) & In-depth interview        & Zoom            & Spanish           & 6 data workers                                                      \\ \hline
\multirow{6}{2.5cm}{\textbf{MANAGERS}}              & Data processing company (Bulgaria) & Expert interview          & Zoom            & English           & 1 company founder                                                   \\ \cline{2-6}
                                                          & \multirow{2}{5.3cm}{Data processing company (Iraq)}     & \multirow{2}{5.3cm}{Expert interview}          & \multirow{2}{5.3cm}{Zoom}            & \multirow{2}{5.3cm}{English}           & 1 general manager                          \\ \cline{6-6}
                                                          &                                    &                           &                 &                   & 1 program manager                          \\ \cline{2-6}
                                                          & Data processing company (Kenya)    & Expert interview          & Zoom            & English           & 1 country manager                                                   \\ \cline{2-6}
                                                          & \multirow{2}{5.3cm}{Data processing company (India)}    & \multirow{2}{5.3cm}{Expert interview}          & \multirow{2}{5.3cm}{Zoom}            & \multirow{2}{5.3cm}{English}           & 1 director of ML services \\ \cline{6-6}
                                                          &                                    &                           &                 &                   & 1 project manager         \\\hline
\multirow{8}{3.5cm}{\textbf{REQUESTERS}}            & \multirow{4}{5.5cm}{Computer vision company (Germany)}  & \multirow{4}{5.3cm}{Expert interview}          & \multirow{4}{5.3cm}{In-person}       & \multirow{4}{5.3cm}{English}           & 1 data protection officer                  \\ \cline{6-6}
                                                          &                                    &                           &                 &                   & 1 co-founder                               \\ \cline{6-6}
                                                          &                                    &                           &                 &                   & 1 product manager                          \\ \cline{6-6}
                                                          &                                    &                           &                 &                   & 1 CV engineer                              \\   \cline{2-6}
                                                          & Machine learning company (USA)      & Expert interview          & Zoom            & English           & 1 product engineer                                                  \\  \cline{2-6}
                                                          & Machine learning company (Spain)    & Expert interview          & Zoom            & Spanish           & 1 lead engineer                                                     \\  \cline{2-6}
                                                          & \multirow{2}{5.3cm}{Computer vision company (Bulgaria)} & \multirow{2}{5.3cm}{Expert interview}          & \multirow{2}{5.3cm}{Zoom}            & \multirow{2}{5.3cm}{English}           & 1 co-founder              \\  \cline{6-6}
                                                          &                                    &                           &                 &                   & 1 CV engineer     \\   \cline{1-6} \cline{1-6}                                 
\end{tabular}%
}
\end{table}
\hspace{1cm}

\subsubsection{Observations}
\hspace{1cm}

Through fieldwork at the BOP and the platforms, we were able to observe interactions among data workers and between them and clients using in-person observation in the Argentinian case and digital ethnography in the Venezuelan one \cite{Horst2012a}. Furthermore, we observed workers’ interactions with crowdsourcing platforms and the software interfaces used to complete annotation tasks. Special attention was paid to the interaction of workers with task instructions. It is important to mention that the instruction documents underwent a twofold form of analysis: On the one hand we analyzed instruction documents as texts through discourse analysis as described in Section 3.3.2. On the other hand, we used the observations conducted to analyze these documents as artifacts or materializations of the data-production dispositif. For the latter form of analysis, the focus was set on documents’ function, provenance, and the interactions they allow or constrain.

The level of involvement regarding observations varied from shadowing to active participant observations. In some cases, we had the opportunity to observe and try the interfaces and perform data annotation tasks for several hours. All observations were recorded as jottings taken in real-time. Those jottings comprised descriptions of briefings, meetings, tasks, documents, communication channels, and interfaces, as well as their advantages and limitations. In parallel, reflections on the researchers’ impressions and perceptions, including explicitly subjective interpretations, were noted.  Simple sketches and, when permitted, photos helped to complete the observations registered. The information gathered in keywords or bullet points was later transformed into complete texts and integrated into more consolidated field notes. In the analysis phase, we combined these field notes with the interview transcripts and coded them following the steps described in the previous subsection.

\section{Findings}
\label{sec:4}

The presentation of our findings comprises a descriptive subsection (\ref{sec:4.1}) and three analytical parts (\ref{sec:4.2}, \ref{sec:4.3}, and \ref{sec:4.4}).

In \ref{sec:4.1}, we present several examples of the different tasks carried out by data workers at each one of our four fieldwork sites. We include details of how task instructions are formulated and communicated to workers and how workers follow or interrogate instructions. Through these descriptions, we seek to locate our analysis in specific settings with specific ways of doing things. 

Next, we move into dissecting the described tasks, instructions, and practices while outlining specific characteristics of the data-production dispositif.  In \ref{sec:4.2}, \ref{sec:4.3}, and \ref{sec:4.4}, we center the findings of the dispositif analysis around our three research questions.  Following RQ1, we describe \emph{discursive practices} such as those involved in the taxonomies used to collect and classify data and the warnings and threats included in them.  Following RQ2, we describe \emph{non-discursive practices and social contexts} such as the obedience to instructions, the dependence of Latin American workers to precarious work, and moments of interrogation and solidarity among workers.  Finally, and following RQ3, we describe some of the \emph{dispositif’s materializations}, such as documents, work interfaces, and tools to measure workers’ performance and surveil them.

\subsection{Different Tasks, Different Instructions}
\label{sec:4.1}

\begin{table}
\caption{Types of tasks in outsourced data work based on Tubaro et al. \cite{Tubaro2020}}
\label{tab:tasks}
\small
\def\arraystretch{1.2}
\resizebox{\textwidth}{!}{%
\begin{tabular}{p{2.5cm}p{3.5cm}p{8cm}}
\cline{1-3}
\cline{1-3}
\multicolumn{1}{c}{\textbf{Task Type}} & \multicolumn{1}{c}{\textbf{Description}}                                   & \multicolumn{1}{c}{\textbf{Examples (based on our fieldwork)}}                                                                                                         \\ \hline
Data Generation                 & The collection of data from the worker’s environment               & “You can earn \$2.5 by completing the task ‘Do you wear glasses?’ Upload a picture of a document with your prescription values now.” \\ \hline
Data Annotation                 & The classification of data according to a predefined set of labels & “Based on the text in each task, select one of these three options: Sexually Explicit, Suggestive, Non-Sexual.”                      \\ \hline
Algorithmic Verification        & The evaluation of algorithmic outputs                              & “You’ll be shown two lists of up to eight search suggestions each. Your task is to indicate which list suggestion is better.”        \\ \hline
AI Impersonation                & The impersonation of an artificial agent                           & As the assistant, the “user will initiate the conversation…you need to use the facts to answer the user’s question.”         \\ \cline{1-3} \cline{1-3}
\end{tabular}%
}
\end{table}

Before moving towards answering our three research questions, we will describe in this section the different tasks available in data work and explored in this study. We use the framework proposed by Tubaro et al. \cite{Tubaro2020} to differentiate the tasks (see Table \ref{tab:tasks}). While this framework was initially conceived to analyze digital platform labor, we think it can also be applied to BPOs in the broader field of data work due to the similar types of tasks available.

Tubaro et al. define three moments in outsourced AI production: “artificial intelligence preparation,” “artificial intelligence verification,” and “artificial intelligence impersonation.” The authors divide AI preparation into the collection of data and its annotation. AI verification involves the evaluation of algorithmic outputs. Finally, AI impersonation, often seen in the corporate and AI-as-a-service sector \cite{Newlands2021}, refers to the non-disclosed “‘human-in-the-loop’ principle that makes workers hardly distinguishable from algorithms” \cite{Tubaro2020}. 

\subsubsection{Data Generation}
\hspace{1cm}

Platform workers are directed to collect data from websites or to produce media content (e.g., text, images, audio, and video) from their devices. For example, a task on Clickrating instructed workers to find information online from companies in the United States, including their address and telephone number. Workerhub required workers to take photos of themselves in certain poses or pictures of family members (including children) and enter attributes of the subjects in these images, including their age and gender. While tasks involving data collection from the web were paid a few cents per assignment, those that involved capturing photos, video, and audio were compensated with a few dollars per file generated. Interviewed workers found the latter type of task attractive because they were the best remunerated. In this case, financial need overcame any privacy concern.

At the BPO, one of the data generation projects produced an image dataset to train a computer vision algorithm capable of identifying fake ID documents. For this purpose, workers were instructed to use their IDs and those of their family members. They took several pictures of the documents and used some of those pictures to create different variations of the ID document (changing the name, the address, the headshot). This way, they produced imagery of authentic as well as fake IDs. The requester of this task was a large e-commerce corporation and Alamo’s most important client. The task is just one of many data-related projects Alamo has conducted for this corporation in the last four years. Because of their ongoing service relationship, the client invested considerable time and money in training Alamo’s workers for each project. One particular characteristic of this relationship is that instruction documents are not created by the client unilaterally but co-drafted with Alamo’s project managers and team leaders. Managers and leaders then serve as support to answer data workers’ questions, should they arise after reading the instructions or while completing the tasks. Because of the sensitivity of the data involved in the ID project, a special interface with several security measures was created to work on the images and store them. Furthermore, all workers had to sign a consent form, allowing the use of their ID document. The request to use their ID and those of family members caused some unease and raised several questions among workers, as one of Alamo’s managers reported. Then, it was the managers’ job to “convince them that their IDs were not going to be used for anything bad, no crime or something. And to do that without revealing too much of the client’s idea because we had signed an NDA.”

\subsubsection{Data Annotation}
\hspace{1cm}

The most common task in data work is data annotation. Projects of this kind were available in all of the studied platforms and the BPO. 

Tasksource and Workerhub provided many image segmentation and classification tasks as platforms specializing in computer vision projects. Some of these tasks included the classification and labeling of images of people according to gender, race, or age categories. Since segmentation tasks required around one hour depending on the size of the image and the number of labels, these tasks are paid more than classifying entire pictures according to a set of categories. Most of the segmentation tasks were designed for training self-driving cars and devices that are part of the internet of things. At the same time, image classification has multiple uses, including content moderation (notably for hate speech and sexual content), healthcare, facial recognition, retail, and marketing. For instance, the same BPO workers that collected pictures of ID documents were subsequently asked to classify and label them as “authentic” or “fake.” In addition, workers had to segment the “fake” ID documents and mark the part of the image they had modified. Finally, they had to annotate the type of modification the image had undergone (e.g., “address has been modified” or “headshot is fake”).

Outside of computer vision, workers of Workerhub were asked to identify hate speech and sexual content in text, notably for social media. For example, in an assignment titled “Identify Racism,” workers were asked to read social media posts and identify whether or not the content included racism or if this judgment was not possible. In another task titled “\$exxybabe69,” workers had to judge if usernames included examples of child exploitation, general sexual content, or none of the previous categories. Video and audio annotation were also present in this platform. For example, in the task “How dirty is this,” workers had to identify if there was sexual or “racy” content in different media types, including audio files and videos. In section 4.2, we will revisit some of these tasks to describe how workers navigated the different taxonomies they encountered to perform data annotation tasks.

\subsubsection{Algorithmic Verification}
\hspace{1cm}

Algorithmic verification involves the assessment of algorithmic outputs by workers. This type of task was observed primarily on the annotation platform Tasker and accessed through Clickrating. Tasker is the internal platform of a major technology company that develops a search engine. They use Clickrating to recruit their workers who, depending on the project, have to sign special contracts and non-disclosure agreements. Algorithmic verification tasks, for example, include assessing how the search engine has responded to a user query, the objects that accompany a search result (e.g., images, maps, addresses), or whether the search result contains adult content or not. In many cases, these assessments include comparing search results with a competitor search engine and assessing which one is more accurate and substantial.

In another example of algorithmic verification, one of the tasks conducted by the BPO Alamo for its largest client (the same e-commerce corporation behind the “ID project”) consisted in verifying the outputs of a model used by the client to moderate user-generated content in their marketplaces. In this case, the task consisted of reviewing content flagged as inappropriate by an algorithm and confirming or correcting the output. For this purpose, the client had provided handbooks that contained each marketplace’s terms and conditions and examples of the specific forms a violation could take. For workers, this task often involved being exposed to disturbing images and violent language, which several interview partners described as “tough.”

\subsubsection{AI Impersonation}
\hspace{1cm}

Impersonation is the rarest type of task, and it was only observed once in the platform Clickrating. Tubaro et al. \cite{Tubaro2020} describe it as a task that occurs “whenever an algorithm cannot autonomously bring an activity to completion, it hands control over to a human operator.” The task that we encountered, developed by a major social media company, asked workers to dialogue with users and respond to their queries according to a set of predefined “facts, history, and characteristics.” If the worker couldn’t answer the user’s query, they were asked to say, “Sorry, I don’t know about that, but can I tell you about...” and then they had to “insert fact related that may be of interest to user.” The platform instructed workers to complete dialogues in the least amount of time, and they have to be logged into the platform in specific 3-hour sessions.

\subsection{Linguistically-Performed Elements}
\label{sec:4.2}
\emph{RQ1: What discourses are present in task instructions provided to outsourced data workers?}

To explore our first research question, we analyze the task instructions \emph{as text}.   Our analysis focuses on the categories and classes used for collecting, sorting, and labeling data as contained in the task instructions. We describe three recurrent elements: (1) the normalization of conventions from the Global North and oriented towards profit maximization, (2) the use of binary classifications and the inclusion of residual categories such as “other” or “ambiguous,” and (3) the discursive elements that aim at constraining data workers’ agency in the performance of data-related tasks.

\subsubsection{Normalized Classifications}
\hspace{1cm}

Taxonomies are the main component of task instructions for data work. They consist of classification systems comprising categories and classes used to collect, sort, and label data. Definitions, examples, and counterexamples usually accompany taxonomies. The number of classes depends on the task and varies according to each platform or company. For instance, assignments on the platform Workerhub usually present a smaller set of labels, ranging from two to a dozen maximum. At the same time, jobs in Tasksource usually feature dozens of classes classified in several categories (e.g., for the semantic segmentation of a road, the category “cars” included labels like “police car” or “ambulance”). The taxonomies that we observed in instructions carry self-evident meanings to the clients but are not necessarily relevant to the annotators or communities affected by the ML system. For instance, the label “foreign language” refers to languages other than English,  and the category “mail truck” only comprises examples of USPS vehicles.

One of the projects conducted at the BPO Alamo consisted in analyzing video footage of a road. A particularity of this project is that the requester did not predefine the labels, but workers were asked to come up with mutually exclusive classes to label vehicles in Spanish. One of the main difficulties of the project, however, lay in the nuances of the Spanish language: Probably oriented towards targeting a broader market, the requester wanted the labels to be formulated in \emph{español neutro} (“neutral Spanish”), i.e., without the idioms that characterize how Argentines and most Alamo workers speak. This contrast led to many instances of discussion among workers and managers about which vehicles’ designations the client would consider “neutral.” 

The mismatch between the classifications that requesters and outsourced data workers consider self-evident becomes critical in cases of social classification. For instance, in the task shown in Example 1, workers were asked to label individuals’ faces for facial recognition according to predefined racial groups that included “White, African American, Latinx or Hispanic, Asian, Indian, Ambiguous,” where the last category should be selected “ONLY if you cannot identify the RACE of the person in the image.” 

\begin{tcolorbox}[standard jigsaw, title=\small Example 1, opacityback=0, colbacktitle=white, coltitle=black]
\small
In this task you will be determining the \textbf{race} of the persons in the images.

You should select only \textbf{one} of the following categories:
\begin{itemize}
    \item White
    \item African American
    \item Latinx or Hispanic
    \item Asian
    \item Indian
    \item Ambiguous
\end{itemize}
\end{tcolorbox}

Beyond the already problematic situation of being asked to “guess someone’s race,” for many interviewed Latin American workers, this type of classification did not make sense, as it was conceived by a US company with a US-centric conception of racial classification, i.e., with little regard for the cultural and racial complexities in Latin America. Similarly, Gonzalo, a data worker for Workerhub, declared having trouble with a task that asked him to label “hateful speech” in social media posts:

\begin{displayquote}
They give you many examples of what they consider “hateful” and not. But, once you’re doing the task, you don’t encounter basic examples, and it’s up to you as a worker to interpret the context and decide what counts as “hate.” Clearly, [the requesters] have their parameters and, if they don’t consider something hateful, they will mark your work as wrong. … For example, a sentence like “kick the latinos out” would not be regarded as “hateful,” and you will not interpret it the same way as a Latino.
\end{displayquote}

These examples of social classification and conceptualization are not just about cultural differences between requesters and data workers, but they reflect the prevalence of worldviews dictated by requesters and considered self-evident to them. Furthermore, the taxonomies respond to the commercial application of the product that will be trained on the data that outsourced workers produce. For example, the instructions to categorize individuals for facial recognition technologies in Example 1 are based on a US-centric definition of “protected group.” Such definitions suggest requesters’ efforts to prioritize the mitigation of legal risks, neglecting the safety of users from social minorities and groups facing discrimination in other contexts, such as ethnic groups with defined caste systems or social categories not protected by US law, such as economic class. 

In another example that shows the inscription of requesters’ profit orientation, workers of a major social media app working through Clickrating, were instructed to evaluate if a post was “building awareness.” Here, “awareness” referred to posts with commercial content and discarded anything personal or political in nature (see Example 2).

\begin{tcolorbox}[standard jigsaw, title=\small Example 2, opacityback=0, colbacktitle=white, coltitle=black]
\small
\textbf{Building awareness} means that the purpose of the [post] is to give information about a brand, product, or other [sic] object. For example, “@Restaurant is awesome for karaoke and the food is delicious!” is building awareness. 
A story is not building awareness if it’s primarily about the author’s life. A story that mentions the name of a business or service without providing much additional context, or a story that refers to a product in passing while its author is sharing one of their experiences, is not building awareness. For example, if the author says, “I’m eating in @restaurant”, the story is not building awareness for the restaurant. 
\end{tcolorbox}

Requesters’ profit orientation is implicit in task instructions and gets inscribed in the ways data workers approach their tasks. For instance, when asked what the procedure would be if they were unsure about annotation instructions, most of the workers at the BPO answered that they would abide by the requesters’ opinion because “their interpretation is usually the one that makes more sense as they know exactly what kind of system they are developing and how they plan to commercialize it.”

\subsubsection{Binary and Residual Categories}
\hspace{1cm}

While complex taxonomies are common in the BPO, the most common classification tasks in platforms are binary. For example, for tasks described as “Does this link title use sensationalist phrasing or tactics?,” “Who is the author of this story?,” and “You will be looking at a bounding box and choosing whether it is around a primary face,” the possible labels were “yes or no,” “human or non-human,” and “yes or no” respectively. As mentioned above, many tasks, especially in Workerhub, are based on categories protected by the United States legislation when defining what counts as hate speech, racism, and other forms of discrimination (See Example 3). However, the classification of whether a text contains hate speech is often reduced to a binary decision without considering the context.  

\begin{tcolorbox}[standard jigsaw, title=\small Example 3, opacityback=0, colbacktitle=white, coltitle=black]
\small
In this task, you will be identifying messages that contain \textbf{hate speech}.

Based on the text, you must select:

\begin{itemize}
    \item \textbf{Hate Speech:} if the username contains hateful content
    \item \textbf{None:} if there is NO hateful or abusive language in the given [sic]
\end{itemize}

\textbf{Definition:}

Select \textbf{Hate Speech} if the text contains \textbf{any} of the following:

\begin{itemize}
    \item Discrimination, disparagement, negativity, or violence against a person or group based on a \textbf{protected attribute}
    \item References to hate groups, or groups that attack people based on a \textbf{protected attribute }
\end{itemize}

\end{tcolorbox}

This form of binary classification often ignores ambiguity or uncertainty, such as when workers are confronted with contexts that are ambiguous or separate from their cultural setting. Moreover, given the impossibility of platform workers to send feedback to requesters, many omissions (involuntary or not) remain unquestioned.  For instance, a task on Workerhub asked workers to classify images as “racist” or not; we observed an image representing several copies of the “crying Wojak” meme wearing kippahs inside a heating oven while the meme “Pepe the Frog” is watching outside. The text above reads: “Changed the wooden doors today frens [sic], this is actually working as intended now!” While this image was, for us, a clear example of antisemitism, a form of racism condemned worldwide, including by the United Nations \cite{UnitedNationsGeneralAssembly1999}, requesters instructed workers not to consider this type of hateful content as “racist.”

In several of the cases where binary categories were prescribed, we encountered a third label, usually called “other” or “ambiguous,” not to designate an additional class that would break the binary classification but to merge errors or instances where the worker cannot apply one of the two labels (see Example 4). Workers are also encouraged to ignore ambiguity altogether. Some Clickrating tasks acknowledge the limits of this binary classification and urge workers to ignore other possible attributes when categorizing data. For example, in an assignment where workers tagged queries for a search engine, the instructions referred to “Queries with Multiple Meanings” for “queries [that] have more than one meaning. For example, the query [apple], in the United States might refer to the computer brand, the fruit, or the music company.” In this case, the task instructed workers to “keep in mind only the dominant or common interpretations unless there is a compelling reason to consider a minor interpretation.” The “dominant” or “common” interpretation of a term in the US may be different to the one in Latin America. Still, we repeatedly encountered similar instructions to deal with instances of ambiguity in many tasks.

\begin{tcolorbox}[standard jigsaw, title=\small Example 4, opacityback=0, colbacktitle=white, coltitle=black]
\small
\textbf{Overview}

Select:

\begin{itemize}
    \item \textbf{Male}: if the boxed face is a male
    \item \textbf{Female}: if the boxed face is a female
    \item \textbf{Other}: if there is no face in the box
\end{itemize}
\end{tcolorbox}

Very often, instruction documents were dated, and requesters provided several updates. For instance, in the task “How dirty is this image/video” on the platform Workerhub, workers were initially instructed to label adult content in images. Later on, the requesters provided videos without updating the documents, confusing some workers because they could not apply the original instructions to the video footage. As a result, the requesters had to update the instructions to include details about video annotation. In another example, a document that asked workers to classify elements in a road revised its definition of “Pedestrians sitting on the ground” to include also those “laying [sic] on benches and laying [sic] or sitting on the ground” (see Example 5). This example also shows that the “other” category that we described above does not represent “everything else” included in the classification but that there are elements deliberately or accidentally left out. In this latter example, the requesters failed to see that people in the streets are not necessarily always “walking” or “sitting,” but that a segment of the population lies or sleeps on them.

\begin{tcolorbox}[standard jigsaw, title=\small Example 5, opacityback=0, colbacktitle=white, coltitle=black]
\small
\textbf{Pedestrians sitting on the ground}

Use the “pedestrian” label if a pedestrian is sitting on the ground, bench, ledge, then use the “pedestrian” label. 

\textbf{UPDATE!!}

Use the “\textbf{PEDESTRIAN LYING DOWN}” label for pedestrians laying [sic] on benches and laying [sic] or sitting on the ground. 
\end{tcolorbox}

\subsubsection{Warnings}
\hspace{1cm}

As we will argue in the following sections, the influence and preferences of powerful actors in data work are stabilized through narrow task instructions, specially tailored work interfaces, managers and quality assurance analysts in BPOs, and algorithms in crowdsourcing platforms.  Some of these processes are part of the tacit knowledge workers have about their position (i.e., \emph{it goes without saying} that workers must carry out tasks according to the preferences of requesters). However, task instructions often make explicit reference to the power differentials between workers and requesters as they include threatening warnings such as “low quality responses will be banned and not paid” or “accurate responses are required. Otherwise you will be banned” (see Examples 6–8). 

Platform workers risk being banned and even expelled from the platform if they fail to follow task instructions. At the BPO Alamo, the communication of instructions is mediated by project managers and team leaders. For this reason, such warnings are not explicit in written documents but are present in reviewing instances and evaluating workers’ performance. Here, any concerns expressed at the workers’ end are filtered through hierarchical managerial structures and hardly ever reach requesters.

\begin{tcolorbox}[standard jigsaw, title=\small Example 6, opacityback=0, colbacktitle=white, coltitle=black]
\small
This is a \textbf{high paying job}, a special job, but to gain access to it and to keep access to it after passing the qualification test, we require patience and \textbf{VERY careful [sic] thought out and accurate responses}.

\textit{Otherwise, you will, unfortunately be banned from the job :(}

\end{tcolorbox}

\begin{tcolorbox}[standard jigsaw, title=\small Example 7, opacityback=0, colbacktitle=white, coltitle=black]
\small

The picture above shows SHIFTING DATA which means that the LiDAR points for stationary objects move or slide around throughout the scene.

\textbf{Any [Project] Tasks with shifting data are not usable by the customer \& have to be cancelled. You will NOT get paid for working on task [sic] with shifting data!!!!!!!!!!}

Every time you get a [Project] Task (before you start working) always turn on dense APC and look around the entire scene to check for shifting data. You will be able to tell that the shift is big enough to be cancelled if it makes any object 0.3m plus larger than it’s [sic] normal size (or if it makes a flat wall 0.3m plus thick) and effects [sic] multiple cuboids.
\end{tcolorbox}

\begin{tcolorbox}[standard jigsaw, title=\small Example 8, opacityback=0, colbacktitle=white, coltitle=black]
\small

We value your individual opinion and review each result, so please provide us with your best work possible. We understand that this can be a tiring task, so if you are in any way unable to perform your best work, please stop and come back once you are refreshed.
You may also see multiple queries with the same kind of visual treatment. 

Please keep your judgments consistent UNLESS you feel that there is some difference in the two that would result in a change of overall score.

\textbf{Judges providing low quality responses will be banned and not paid.}

\end{tcolorbox}

These messages encode a precise definition of “accurate responses”: Accuracy is classified according to what the client believes to be an accurate truth value, while divergence from that value is considered inaccurate. Here, too, the classifications that make sense to requesters have prevalence. This is why workers at the Argentine BPO are permanently encouraged by management to think in terms of “what the client might want and what would bring more value to them.”

Given the social and economic contexts in which the outsourcing of data work occurs, warnings and threats of being banned or fired reinforce the hierarchical structure of the annotation process and compel workers to follow the norms as instructed or risk losing their livelihood. In the next section, we will present evidence of how the social contexts of workers and the fear of losing their job shapes how assignments are carried out.

\subsection{Non-Linguistic Practices and Social Context}
\label{sec:4.3}

\emph{RQ2: How do outsourced data workers, managers, and requesters interact with each other and instructions to produce data?}

This section explores our second research question. Here, we focus on analyzing interviews to describe the contexts in which data-related tasks are carried out and the interactions they enable. We describe (1) the social contexts of Latin American workers that lead to their dependence on data work regardless of the labor conditions, (2) the elements that contribute to the unquestioning obedience to instructions, and (3) moments of subversion of rules as well as workers’ organization and solidarity. 

\subsubsection{Poverty and Dependence}
\hspace{1cm}

Being an impact sourcing company, Alamo employs around 400 young workers who live in slums in and around Buenos Aires. As stated on its website, Alamo specifically recruits workers from poor areas as part of its mission. As Natalia, one of the BPO’s project managers, describes, this is a population that does not receive many opportunities in the Argentine labor market:

\begin{displayquote}
They are very young, and a bit, you know… Alamo works with people another company wouldn’t hire, so people who live in areas… slums with difficulties, with a very low socioeconomic level. That’s something the company pays attention to when it comes to recruiting, and if during the interview we detect that the candidate could have an opportunity somewhere else, we prefer not to hire that person and hire someone else.
\end{displayquote}

One particularity of Alamo is that it provides workers with a regular part- or full-time salary. This form of employment contracts with the widespread piece-wage model in platforms.  The salary Alamo’s workers received in 2019 was the equivalent of 1.70 US dollars per hour which was the minimum legal wage in Argentina. Despite the low wages and exhausting tasks such as semantic segmentation or content moderation, all interviewees were satisfied as the company offers better conditions than previous experiences.  According to a report published in November 2020 by the national Ministry of Production \cite{schteingart2020}, the unemployment rate in Argentina is 10\%, and 35\% of the employed labor force is not registered. Argentina has a long tradition of undeclared labor. This way, employers avoid paying taxes while workers remain without protection or benefits. 

Behind the numbers are people like Nati, who did different types of precarious work before working for Alamo. She started at the BPO as a data annotator and quickly became a reviewer until being offered a position as an analyst in the company’s QA department.  Like other Alamo workers, she acknowledges the difficulty of securing a desk job somewhere else. Moreover, many of our research participants mentioned being proud of the work they do at Alamo because a desk job has “a different status.” For several of them, working at Alamo means finally having a steady income and breaking with generations of informal gigs, for example, in the cleaning or construction sectors. As Nati explains, what Alamo offers is better than the alternatives:

\begin{displayquote}
That was the situation at home; we were going through a rough time.  My mother was out of work because her former boss had found someone else to clean, and I had lost my job too.  So I needed a job and when I found this one I was surprised to work at a friendly place for a change! Now I have a desk, a future, and I feel appreciated. This is new to me.
\end{displayquote}

In the case of the platforms, they have thrived in the Venezuelan economy, which is characterized by the highest levels of inflation in the world, with an average of 3,000\% in 2020 \cite{AgenceFrance-Presse2021}. All participants that we interviewed from Venezuela stated that the “situación país” [country’s situation] was the main reason they resorted to online work. Workers reported difficulty finding employment in the local labor market, especially for income that is not dependent on the national currency, the bolivar, which devalues quickly. For example, Rodrigo, a Clickrating worker, quit his job as an information technology consultant because online platforms were the only way he could earn US dollars. He explains the monetary situation of his country as follows:

\begin{displayquote}
There are two types of currency exchange rates: the official rate dictated by the government and the one used on the black market, which everyone uses. Everyone knows this black-market exchange rate. It’s an Instagram profile that posts the average exchange rates of several independent currency exchange websites. They make this average and post the fluctuation several times per day, which is the exchange rate that we use today.
\end{displayquote}

Platforms’ low entry barriers make outsourced data work an attractive\,---\,and sometimes the only\,---\,source of employment during social, economic, and political crises. Data workers earn 15 to 60 US dollars per week, the average being around 20 dollars, which is substantially higher than the minimum wage in Venezuela reported by workers to be around 1 US dollar per month in March 2021. 

Dependency on the platforms is exacerbated by the high unemployment levels and reduced government support during the COVID-19 pandemic. In this situation, workers have limited access to subsidies and pay for services such as healthcare from their income \cite{Posada2022}. For example, Olivia, one of the Tasksource workers, was diagnosed with diabetes and has to self-fund the costs of insulin. For her, losing access to the platform is a life-threatening situation:

\begin{displayquote}
Imagine, with this pandemic, what can I do? My medical situation does not allow me to go outside and risk getting the [coronavirus disease]. If I get it, I’ll die. For this reason, I cannot take the risk and expose myself to something worse; I can’t risk this job either because this is my only source of income.
\end{displayquote}

This dependency affects the labor process as well. Workers usually do not choose which tasks to perform, even if they disagree ethically with their assignments. When asked what criteria Carolina, a Clickrating worker, uses to choose a task over another, she answered:

\begin{displayquote}
My priority is to get the tasks that pay the best. But I don’t even have that choice. The platform restricts which jobs are available here in Venezuela, so I have to make the most of it to earn the minimum and get paid as soon as I get one task.
\end{displayquote}

By “minimum,” Carolina refers to the minimum income workers can transfer out of the platforms, which is another form of creating dependency. Platforms establish a minimum of 5 to 12 US dollars before they make payments and, if a worker cannot achieve this threshold, they have to wait for a week before withdrawing their salary. This payment process is a form of institutionalized wage theft implemented by the platforms. Workers lose the money they have worked for if they get banned before reaching the threshold for payment.  

\subsubsection{Obedience to Instructions}
\hspace{1cm}

In its outsourcing capacities, Alamo focuses on data-related services ordered mainly by machine learning companies.  Even if they display some similarities, each of those projects is different from the previous ones, and workers need to be briefed regularly. Depending on the difficulty and extension of the task, briefings can be more or less sophisticated and involve more or fewer actors, meetings, and processes. Sometimes, the instructions for new projects are sent by the requester via email in a PDF document. One of the area managers receives that information and transmits it to a project manager, who would then put together a team and work closely with their leader. Depending on the degree of difficulty, one or more meetings with the team will be held to explain the project, answer questions, and supervise the first steps.  When handling large projects from multinational organizations, Alamo invests a considerable amount of resources in the briefings. No matter how big or small the requester, briefings at Alamo consist of getting the workers acquainted with the expectations of the requesters and are a way of making sure workers are on the same page and thinking similarly:

\begin{displayquote}
[Quality assurance analyst with Alamo] The information from the client usually reaches the team leader or the project manager first, and, at that moment, what we do is to have a meeting for criteria alignment… that is generally what we do.  The team meets to touch base and see that we all think in the same way.
\end{displayquote}

These briefings give workers a framework for new projects and are instrumentalized by the company as the first instance of control, aiming at reducing room for subjectivity. Further control instances, aiming to ensure that data work is done uniformly and according to requesters’ expectations, take place in numerous iterations where reviewers and team leaders review and revise data and go back to the instruction documents or contact the requester to clarify inconsistencies. 

In companies like Alamo, \textit{data quality} means producing data precisely according to the requester’s expectations. According to Eva, a BPO manager in Bulgaria, this view on data quality is commonplace in data services companies. In the following excerpt, she summarizes the importance and main function of instructions and further instances of control, i.e., making sure that the workers interpret the data homogeneously:

\begin{displayquote}
Normally, issues in data labeling do not come so much from being lazy or not doing your work that well. They come from a lack of understanding of the specific requirements for the task or maybe different interpretations because a lot of the things, two people can interpret differently, so it’s very important to share consistency and, like, having everyone understand the images or the data in the same way.
\end{displayquote}

The interviews we conducted with requesters show that the priority behind the formulation of task instructions is producing data that fits the requester’s machine learning product and the business plan envisioned for that product. What does not match the requester’s instructions is considered low-quality data or “noise.” Dean is a machine learning engineer working for a computer vision company in Germany. He reported on this widespread view as follows: 

\begin{displayquote}
Dean: Noise is what doesn’t fit your guidelines. 

Interviewer: And where do those guidelines come from? 

Dean: We say, “actually we want to do this, we want to do that,” and then, of course, since the client is the king, we translate that business requirement into something like… into a requirement in terms of labels, what kind of data we need.
\end{displayquote}

As described in Section 4.2.3, compliance with requesters’ views is made explicit in instruction documents in the form of warnings for workers. Those documents are usually the only source of information and training platform workers have to complete their tasks. However, the case of the platform Tasksource is slightly different, as it employs Latin American coaches to brief and explain to workers how to interpret instructions and annotate tasks.  This approach is similar to the one used by the BPO Alamo and described above.  However, at Tasksource, briefings take place in week-long unpaid digital courses called “boot camps” and later evaluation periods called “in-house.” 


The use of the military and correctional term “boot camp” could be interpreted as reflecting this training’s purpose: conditioning workers to obey tasks without question. Ironically, even though the platform employed workers to help train artificial agents, they were supposed to behave like “robots,” according to a Tasksource worker named Cecilia:

\begin{displayquote}
When you start, they tell you: “To be successful in this job, you have to think like a machine and not like a human.” After that, they explain to you why it has to be like that. For example, you are teaching a [self-driving] car how it has to behave. When you segment an image, there is a police car, and you label it like a regular car, the [self-driving] car will think it’s a regular car and, if it crashes against it, something terrible can happen. The mistake was not of the car that crashed into the police vehicle, but it’s yours as a tasker, as a worker, who taught the car to behave like that.
\end{displayquote}

Platform workers serve a similar role as BPO employees in reinforcing the primacy of instructions and requester intent to complete tasks effectively, producing data that fits model and revenue plan while shifting the responsibility for failures on workers. In this context, obedience to instructions is critical for data workers to keep their job and make a living. The fear of being fired, banned from the job, or not being paid for the task reinforces the disposition of workers to being compliant, even when instructions look arbitrary. This is what Rodolfo, Tasksource worker, reported:

\begin{displayquote}
That is why I don’t like that platform very much. Because they give us the instructions and we have to follow. And there are many cases where, if you don’t complete the task really to perfection, according to what they want or what they think is right, they just expel you. Just like that, even if you followed the instructions thoroughly. 
\end{displayquote}

\subsubsection{Worker Solidarity and Organization}
\hspace{1cm}

Not everything is imposition and obedience in data work.  There are also several expressions of workers organizing to improve working conditions and help each other deal with tasks and make the most out of them. 

For instance, Alamo’s employment model that includes data workers as part of the company’s permanent staff instead of having them as contractors results from workers organizing to demand receiving a fixed salary and benefits. As reported by one of Alamo’s reviewers, Elisabeth, in 2019, further workers’ demands were being negotiated with the company:

\begin{displayquote}
We asked for a couple of things like the possibility of home office and a better healthcare plan. We are organizing many things. It’s being negotiated.
\end{displayquote}

In 2020, probably also motivated by the Covid-19 pandemic, Alamo’s data workers were finally allowed to work remotely.  It is worth mentioning that before 2020, every other company’s department and management were allowed to work at least some days of the week remotely while the data workers could not. 

In the case of the platforms, data workers organize in virtual groups and fora. The existence of virtual and local groups of workers that provide solidarity and support has been reported in other examples of platform labor \cite{Wood2018d, Delfanti2019a, Qadri2020}. In the case of Venezuelan data workers, we observed similar situations. Because we used convenience and snowball sampling and worker groups on social media as a starting point, all the interviewed participants were directly associated with them. Participants use these independent and worker-led spaces to exchange information about which tasks pay more and are less challenging to complete and warn each other about non-reliable requesters. One of the aspects that workers paid significant attention to was the presence of bugs in the tasks. When asked about their existence, Yolima, a worker with the platform Tasksource, said to us:

\begin{displayquote}
Errors occur all the time. But, since we are in groups on Facebook and Whatsapp, we alert each other and say, “Hey, don’t do this task because it has a bug. It will flag you as mistaken even if you have done everything ok.”
\end{displayquote}

Some smaller groups, with high entry barriers to ensure privacy and trustworthiness among members, recommend specific tasks over others. For example, when describing tasks with sexual or violent content, Estefanía, one of Clickrating’s workers, stated:

\begin{displayquote}
I don’t like those tasks with pornographic content. I do them only when my friends from the groups say, “look, this is a good task, here’s the link.” I don’t have to look for good tasks, and that’s great. I just have to log into my account and do the annotation without worrying about which tasks to do.
\end{displayquote}

Some users of these smaller groups also craft guides to explain the instructions to their peers. Most interviewed workers stated that their knowledge of the English language was limited. Since Tasksource and Clickrating only presented instructions in that language, and Workerhub provided automated translations with errors, these guides in Spanish are a fundamental tool for workers. They are written by workers for their peers and contain Spanish translations of taxonomies, definitions, and examples. They also provide further explanations about the contexts in which workers can apply the taxonomies, avoid being banned by the algorithm, and maintain high accuracy scores.  For example, in the introduction of a guide for a task to annotate hate speech in text for Workerhub, a user wrote:

\begin{tcolorbox}[standard jigsaw, title=\small Example 9, opacityback=0, colbacktitle=white, coltitle=black]
\small
IMPORTANT INFORMATION

What I’m sharing in this guide is based on my experience with the task. I’ll try to explain as best as I can the tips that I consider are the most important to avoid being banned and the essential information to understand the task.

BE CAREFUL

The task “No Hatred” is not available on all accounts. You must have been paid AT LEAST ONCE. 

IT’S IMPORTANT THAT YOU CONSIDER THIS GUIDE FOR WHAT IT IS: A “GUIDE” made for you to understand the task better. You must earn real experience by doing the task with perseverance and dedication. 

\end{tcolorbox}

Work practices in the data-production dispositif are not informed exclusively by the relationships between requesters, intermediaries (platform or BPO), and individual workers. They are also dependent on the networks formed by the latter group. This can be observed in BPOs where data workers share the same office space and constantly consult and advise each other on conducting projects more quickly and easily. Among platform workers, online groups help to choose what tasks to carry out, and such decisions are influenced by recommendations and guides from peers who evaluate instructions from requesters.

\subsection{Dispositif’s Materializations}
\label{sec:4.4}

\emph{RQ3: What artifacts support the observance of instructions, and what kind of work they perform?} 

In this section, we focus on the third research question. Based on the observations conducted at the crowdsourcing platforms and the BPO company, we present three of the many possible materializations of the data-production dispositif: (1) the function of diverse types of documents that embody the dispositif’s discourses, (2) the platforms and interfaces that guide and constrain data work, and (3) the tools used by managers and platforms to surveil workers and quantify their performance.

\subsubsection{Documents as Artifacts}
\hspace{1cm}

In Section \ref{sec:4.2}, we focused on the \emph{content} of instruction documents to describe the discursive elements comprised in them. Here, instead, we look into a variety of documents\,---\,instructions included\,---\,to analyze them as \emph{artifacts}, focusing on their form, function, and type of work they perform.

One common document related to data work at BPOs is that containing metadata and project documentation.  Alamo, for instance, records the details of each project in several documents that vary in form and purpose according to the task and the requester. Often, that documentation aims to preserve the evolution of task instructions, registering changes requested by clients. Keeping this type of documentation functions as a form of “insurance” for Alamo and can help resolve discrepancies if requesters are not satisfied with the service provided. In those cases, project documentation serves as proof that data was produced as instructed. Documents containing project details can also serve the purpose of preserving situated and contingent knowledge that would otherwise get lost and could help improve future work practices \cite{miceli2020}. Sometimes, these documents become artifacts that cross Alamo’s boundaries and reach the requesters.  For them, the documents might have a factual function (in terms of the information they want or need) or a symbolic one (to reassure clients that Alamo is \emph{at their disposal}). Alamo’s QA analyst Nati describes this as follows:

\begin{displayquote}
We send a monthly report to the clients, including what was done and problems we encountered; we set objectives for the following month and send an overview of the metrics. Some clients don’t even look at the report but insist on receiving it every month.  Others value it and use the information to report to leadership or investors within their organization. 
\end{displayquote}

The documents produced by the BPO are tailored to be valuable for requesters. Conversely, the documents formulated by requesters often remain unintelligible for data workers, even if they are the primary addressees, as in the case of instruction documents. In many cases, language is the main issue hindering the intelligibility of documents: Most of the workers we interviewed have limited knowledge of the English language and reported using translation services, notably Google Translate, to understand the instructions provided by requesters. As mentioned in the previous section, one of the main reasons platform workers resort to guides written by peers is that they are written in Spanish.  But beyond language differences, elements of the taxonomies used in documents can also be confusing, as explained by Tasksource’s worker Yolima and described in Section 4.2.1:

\begin{displayquote}
For the [categories], they are made in the United States, I think. I don’t know what they would call a laundry sink\footnote{We use this term to translate “lavadero,” a commonplace in Latin American homes for the washing of clothes equipped with a washboard basin and sink.}, a shower, or parts of the bathroom. Most of the time, my mistakes were with parts of the bathroom, especially around the shower, the tap, and those things. That was confusing because that was a shower for me, but it was something else for [the platform]. 
\end{displayquote}

The confusion produced by the different languages is not merely a matter of cultural bias. Looking for cheap labor, platforms and requesters target the Venezuelan market but ignore the language barriers and formulate instructions in English. Moreover, further documents that workers encounter in their work, such as privacy policies, contracts, and non-disclosure agreements, are also prepared in English and remain, partially or totally, unintelligible for them. Workers usually sign these documents without understanding the full scope of their contractual relationship with their employers. Along with instructions, these documents embody the data-production dispositif. They are a materialization of normalized discourses and practices that shape data workers to be dependent and, therefore, obedient, while their subjectivities as Spanish-speaking Latin American workers are ignored and erased.

\subsubsection{Work Interfaces}
\hspace{1cm}

In BPOs like Alamo, choices regarding which platform will host the data and will be used as a tool are made by clients. In many cases, the requester has developed an annotation software specifically tailored to the needs of their business and the dataset to be produced. In other projects, the company uses a commercial platform designed by a third party.  In this case, the client would suggest the tool that best fits their needs among several choices available on the market.

The choice of a specific tool comes with limitations that, in one way or another, constrain data workers’ agency to interpret and sort data. The most notorious one is that the taxonomies contained in instruction documents are also embedded in the software interfaces that workers use to collect, organize, segment, and label data. Workers usually interact with a drop-down menu containing all the classes or attributes they are allowed to apply to data. Most interfaces do not allow workers to add further options to the list of pre-defined labels that they receive from requesters. This is most prominent in software interfaces specially designed by requesters and tailored to specific projects. In those cases, the software interfaces that mediate between workers and data are designed to ensure that tasks are completed according to particular parameters pre-defined by requesters and made explicit in the instruction documents. 

In the case of regular data work tools for commercial use, the impossibility of changing the predefined categories or adding more classes is perceived as a limitation that makes data work harder at BPOs and requires communication throughout hierarchical structures until the requester is reached.  Jeff, one of the managers leading a BPO in Iraq, reports on this issue:

\begin{displayquote}
There was a limitation on the annotation tool that they were using. They were relying on an open-source platform that doesn’t have that feature that lets you add or create predefined attributes, which makes the work many times easier.
\end{displayquote}

\begin{figure}[h]
    \centering
    \includegraphics[scale=0.45]{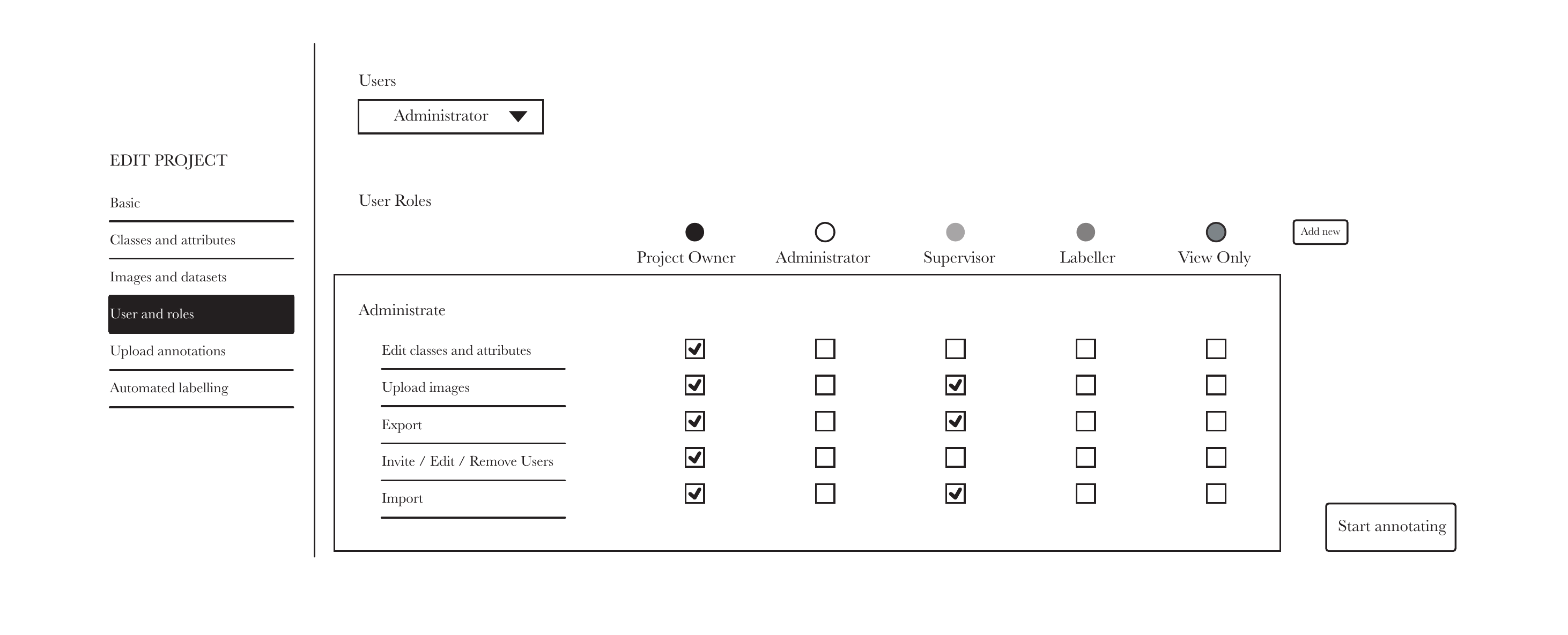}
    \caption{Commercial data annotation tool. Only the project owner can grant rights to data workers.}
    \label{fig:annotation_tool}
\end{figure}
\hspace{1cm}

Some of these generic tools give the project owner\,---\,generally the requester or a BPO’s project manager\,---\,the faculty to allow workers to add further options to the classification system (see Figure \ref{fig:annotation_tool}). However, this does not seem to be a widespread practice at Alamo. Among the many projects that we had the opportunity to observe during fieldwork, only once were data workers allowed to co-create the taxonomy around which data was organized and annotated.

\begin{figure}[h]
    \centering
    \includegraphics[scale=0.40]{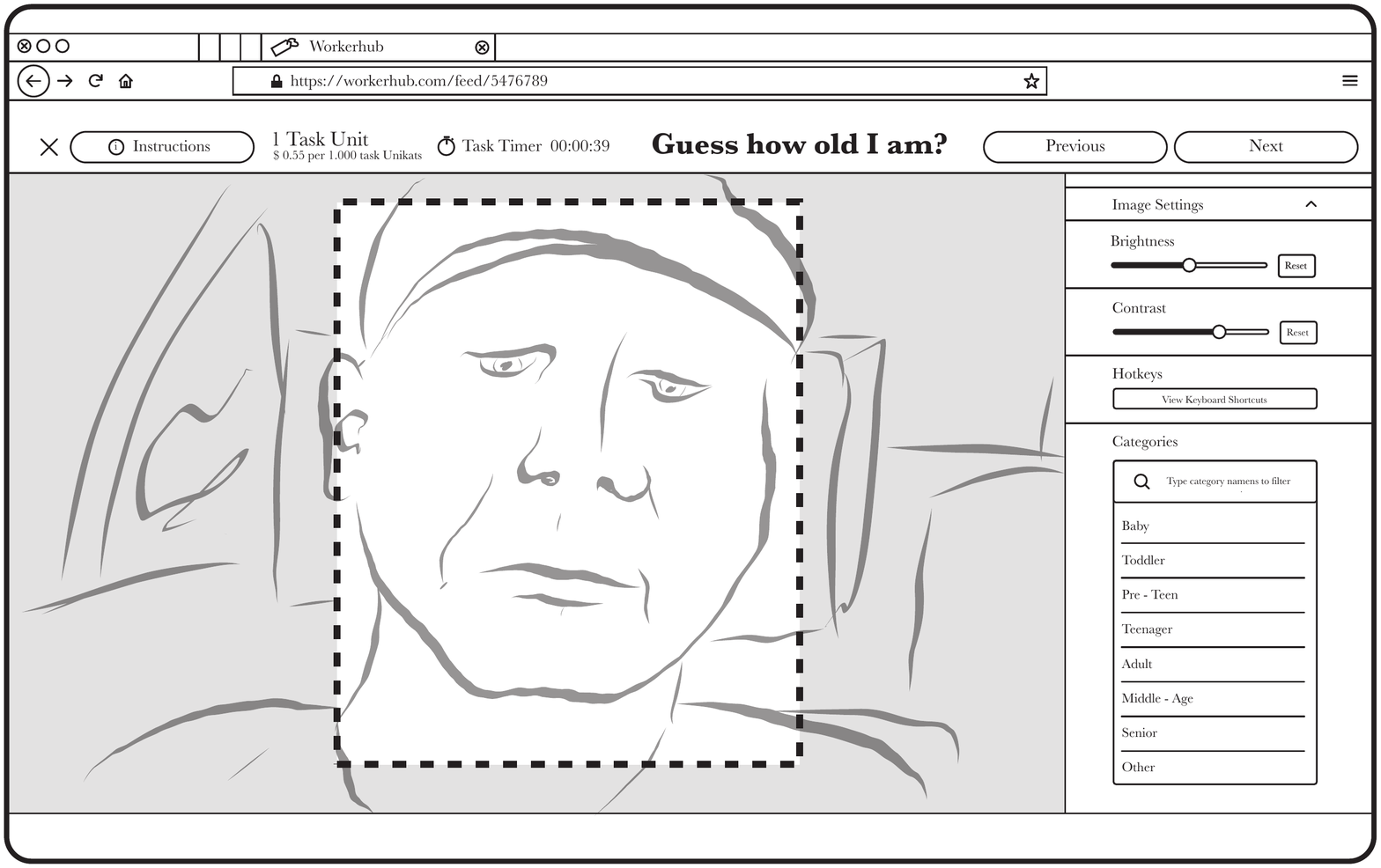}
    \caption{Age-based classification of images on Workerhub's interface.}
    \label{fig:workerhub}
\end{figure}

In the crowdsourcing platforms that we studied, only Clickrating presented external interfaces, meaning that workers had to log into internal annotation platforms of clients, notably in the case of tasks requested by major technology companies. For Tasksource and Workerhub, workers interacted with data annotation interfaces developed by these companies. In both cases, the screen displayed a top bar with an accuracy score or the percentage of tasks submitted by the worker that the platform judged accurate. For Workerhub, the top bar also showed the number of annotations completed for the assignment, the earnings, the time spent per task, and a button to display the instructions (see Figure \ref{fig:workerhub}). On both platforms, the labels were available in the right sidebar alongside tools to zoom in and out and configure the visibility of the data. In all three platforms, workers could not change the predefined labels or suggest changes. 

The interfaces present in the BPO and the platforms feature gamification elements (scores and timing) to speed up the labor process and keep workers focused on the tasks at hand. The over-reliance on speed privileges action over reflection and increases the alienation between workers and the production process. Even tasks that ask for workers’ judgment, such as those present in Clickrating, are timed and reward fast thinking. That said, unlike the other platforms, they offer room for comments to evaluate algorithmic outputs that can be substantial, creating more engagement for workers beyond narrow annotation tasks.

\subsubsection{Tools to Assess Worker Performance}
\hspace{1cm}

To differentiate itself in the very competitive market of outsourced data services, the BPO Alamo makes a selling point out of its performance metrics and quality assurance mechanisms. The company puts much effort into developing more and better ways of measuring performance and quality, and transforming those into numbers and charts the client may perceive as valuable.  As a response to market demands, quality controls intensify, which results in pressure and surveillance for workers. Moreover, the need for quantifiable data to translate “quality” into a percentage exacerbates the standardization of work processes, which, once more, results in less room for workers’ subjectivity.

Alamo has highly standardized processes that include a team leader and several reviewers per team and a quality assurance (QA) department using several metrics to ensure that projects are conducted in accordance with the requesters’ expectations.  In addition, team leaders and the QA department use metrics to quantify workers’ labor. As a token of transparency, sometimes workers’ scores are shared with clients. Noah, one of the BPO’s team leaders, describe the function of metrics within Alamo and concerning its clients:

\begin{displayquote}
We have metrics for everything. They can be individual, for personal output, or they can be general in the project. We have some to measure correct and incorrect output, there we see where we fail, where we can give more support to the team so that those errors are corrected, how we can solve those problems. In QA, what they do is metrics. Metrics, and ensure that the quality provided to the client is high.
\end{displayquote}

In platforms, workers are also constantly evaluated with accuracy and speed metrics.  Instead of being managed by company employees like in the case of Alamo, platform workers are assessed and controlled by algorithms. All platform workers we interviewed reported being often banned from tasks because the algorithms negatively evaluated their performance. Of course, this represents a serious obstacle to maintaining a stable income, especially when it is permanent. This is what Juan, a worker of Workerhub, reported:

\begin{displayquote}
Juan: The platform pays every Tuesday. Once they ban you, you lose all your credits, in the sense that, without an account under your name and email, you can’t open a new account and access the money you’ve earned.

Interviewer: Did they tell you why?

Juan: No. I could have asked in the [platform managed] Discord channel, but if you ask anything, you get banned. They are the ones who command… they are the ones who decide. I was banned without cause because my accuracy was high. I never knew why they expelled me. 

Interviewer: How did you realize you were banned?

Juan: One day, I couldn’t access my account. … I created another account with the same email, worked for a week, and they banned me again. They didn’t pay me. Some of my colleagues from the same neighborhood and cousins who work for the platform told me: “Don’t create an account with the same email because they won’t let you. They will let you open it, but then they won’t pay you.”
\end{displayquote}

The algorithms that assess worker performance in the three platforms that we have studied follow exactly the same three-step process: First, workers have to work with the same data again after some time. For example, if the task involves categorizing photographs of flowers according to their colors, if a worker marks the same image differently, the algorithm will consider it “spam.” The second mechanism is to verify workers’ answers with previously labeled data. If there is a mismatch, the algorithm will assume that the worker is not performing their activities “accurately.” Finally, from interviews with workers and previous observations in Amazon Mechanical Turk \cite{Moreschi2020}, a platform that is not well established in Latin America outside of Brazil and, therefore, not the focus of this study, the third method used by algorithms is to compare workers’ answers with those of peers and assume that the most common answer is the correct one. Many of the workers’ groups that we encounter provide guides, so workers do not diverge from the responses of the majority and, thus, keep high levels of accuracy from the perspective of the algorithms.

\section{Discussion}

In concurrence with previous work \cite{geiger2020,miceli2020, miceli2021a, scheuerman2021}, we have observed that workers collecting, interpreting, sorting, and labeling data do not do so guided solely by their judgment: their work and subjectivities are embedded in large industrial structures and subject to control. Artificial intelligence politics are inextricably connected to the power relations behind data collection and transformation and the working conditions that allow preconceived hegemonic forms of knowledge to be encoded in machine learning algorithms via training datasets. Labor conditions and economic power in the production of ML datasets manifest in decisions related to what is considered data and how each data point is interpreted. 

While task instructions help data workers complete their tasks, they also constitute a fundamental tool to assure the imposition of requesters’ worldviews on datasets. Sometimes, the meanings and classifications comprised in data work instructions appear self-evident to workers, and a shared status quo is reproduced on the dataset. Often, however, the logic encoded in the instructions does not resonate with them. This could be due to cultural differences between requesters and data workers, lack of contextual information about the dataset’s application area, perceived errors that cannot be reported, or simply because the tasks appear ethically questionable to workers. In such cases, another form of normalized discourse persists: that of a hierarchical order where service providers are conditioned to follow orders because “the client is always right” and workers should “be like a machine.” 

According to Foucault, discourse organizes knowledge that structures the constitution of social relations through the collective understanding of the discursive logic and the acceptance of the discourse as a social fact. A normalized discourse is, therefore, what goes without saying. This way, the prevalence of requesters’ views and preferences does not need to be explicitly announced to workers. Instead, such implicit knowledge influences how the outsourced data workers that we observed and interviewed perform their tasks: carefully following instructions, even when they do not make sense to them or when they do not agree with the contents and taxonomies in the documents. The context of poverty and lack of opportunities in the regions where data production is outsourced is also fundamental as it makes workers dependent on requesters and, thus, obedient to instructions.

 Finally, artifacts such as narrow work interfaces with embedded predefined labels, platforms that do not allow workers’ feedback, and metrics to assess workers’ “accuracy” (understood as compliance to requesters’ views) constitute discursive materializations that, at the same time, ensure the perpetuation and normalization of specific discourses.

\textbf{All these elements combined \,---\,the predefined truth values encoded in instructions, the work practices and social positions of workers, and materializations such as interfaces\,---\, constitute the data-production dispositif.} Without any of these elements, the dispositif would not be able to function as such.  As Foucault puts it, dispositifs respond to an “urgent need” \cite{foucault1980} that is historically and geographically contingent. \textbf{The data-production dispositif responds to the voracious demand for more, cheaper, and increasingly differentiated data to feed the growing AI industry} \cite{Bender2021,Crawford2021}.  Its goal is to produce \emph{subjects} that are compliant to that need.

The Foucauldian notion of \textit{subject} has a twofold meaning, with subjects, on the one hand, being producers of discourse and, on the other hand, being created by and subjected to dispositifs. All subjects are entangled in dispositifs and have, therefore, tacit knowledge of \textit{how to do things} within specific contexts. This tacit knowledge includes “knowing one’s place” and what is expected from each subject depending on their position. Thus, data workers know that subjects in their social and professional position are implicitly expected to comply with client’s requests. This way, dispositifs normalize and homogenize the subjectivities of those they dominate, producing power/knowledge relationships that shape the subjects within the dispositif according to certain beliefs, actions, and behaviors that correspond to the dispositif’s purpose \cite{foucault1971, foucault1982}.

Following Foucault’s perspective, we argue that \textbf{the goal of the data-production dispositif is creating a specific type of worker, namely, outsourced data workers who are kept apart from the rest of the machine learning production chain and, therefore, alienated.  Data workers are recruited in impoverished areas of the world, often under the premise of “bringing jobs to marginalized populations,” but are not offered opportunities to rise socially or professionally in terms of salary and education. Data workers who are surveilled, pushed to obey requesters and not question tasks, and who are constantly reminded of the dangers of non-compliance}. Data production cannot be a dignifying type of work if it does not provide workers with a sustainable future.

The implications of this data-production dispositif designed to constrain workers’ subjectivities and perpetuate their \emph{alienation, precarization,} and \emph{control}, will be unpacked in the following subsection.

\subsection{Implications}

As the generous corpus of research literature dedicated to mitigating bias in crowdsourcing suggests, controlling workers’ subjectivities is considered essential to avoid individual prejudices being incorporated in datasets and, subsequently, in machine learning models. 
However, as we have shown with our findings, unilateral views are already present at the requesters’ end in the form of instructions that perpetuate particular worldviews and forms of discrimination that includes racism, sexism, classism, and xenophobia. 

Given its characteristics, the data-production dispositif is detrimental to data workers and the communities affected by machine learning systems trained on data produced under such conditions. 
To close this paper, we would like to make a call to dismantle the dispositif. 
However, before going into the implications of our call, it is crucial to consider that we never cease to act within dispositifs and, by dismantling the data-production dispositif, we would inevitably give rise to another one. 
Therefore, we discuss here ways of dismantling the data-production dispositif \emph{as we know it today}, that is, by changing the material conditions in data work and making its normalized discourses explicit. 

\subsubsection{Fighting Alienation by Making the ML Pipeline Visible to Workers}
\hspace{1cm}

Substantial efforts in research and industry have been directed towards investigating and mitigating worker bias in crowdsourcing. Many of these initiatives portray data workers as bias-carrying hazards whose subjectivities need to be constrained to prevent them “contaminating” data. This widespread discourse within the data-production dispositif gives place to narrow instructions and work interfaces and the impossibility of questioning tasks. Workers are required to “think like a machine” to be successful in the job. Moreover, data workers are often kept in the dark about requesters’ plans and the machine learning models that they help train. Such conditions lead to workers’ alienation as they are kept apart from the rest of the ML production chain.

Researchers have often referred to data workers as data labelers and content moderators, practicing \textit{ghost work} \cite{Gray2019} that remains “invisible” \cite{Roberts2019}. However, as Raval \cite{raval2021} accurately argues, it is worth asking \emph{invisible for whom} and, most importantly, “what does this seeing/knowing—hence generating empathetic affect among Global North users—provide in terms of meaningful paths to action for Global South subjects (workers and others)?” Breaking with the alienation of data workers means much more than rendering them visible. It rather requires making the rest of the machine learning supply chain visible \emph{to them}. It means providing information and education on technical and language matter that could help workers understand how their valuable labor fuels a multi-billion dollar industry. This also concerns questions of labor organization and unionizing: For instance, the recently-created Alphabet Workers Union has taken steps in this direction by including contractors\,---\,many of them outsourced data workers. 
To help counter their alienation, researchers and industry practitioners need to regard data workers as tech workers as much as we do when we think of engineers. 

\emph{Why would requesters want to educate data workers and disclose technical or commercial information to them?}

As mentioned above, the design of the tasks that we encountered failed to acknowledge and rely on the unique ethical and societal understanding of workers to improve the annotations and, with them, models. We found that the BPO model generates a stronger employment relationship with workers compared to platforms, notably Workerhub and Tasksource, which translates into higher engagement with the tasks at hand. Furthermore, BPO workers interviewed by us said they wished they knew more about the requesters’ organizations and products because this would help them understand their work and perform better. In this sense, expanding instructions to include contextual information about the task, its field of application, and examples that show its relevance for systems and users could improve data workers' motivation and satisfaction, and help them understand the value of their labor within ML supply chains.

\subsubsection{Fighting Precarization by Considering Data Workers Assets}
\hspace{1cm}

One of the most pressing ethical and humanitarian concerns surrounding outsourced data work is the workers’ quality of life. The data-production dispositif is designed to access a large and cheap labor pool and profit from workers’ precarious working conditions. It is not a coincidence that, in Latin America, the platforms we encountered were established primarily in Venezuela, a country mired in a deep socio-political crisis exacerbated by the COVID-19 pandemic, and that the BPO company in Argentina recruited its workers from low-income neighborhoods. 

While the arrival of these platforms and BPOs has allowed many workers to circumvent the limits of their local labor markets, the system of economic dependency and exploitation that they reproduce hinders efforts for sustainable development that include access to decent work, and economic growth \cite{UnitedNations2015}. Labor is an often overlooked aspect in discussions of ethical and sustainable artificial intelligence \cite{Posada2021b}. We argue that we cannot truly create fair and equitable machine learning systems if they depend on exploitative labor conditions in data work.

\emph{Why would requesters want to improve labor conditions in outsourced facilities?}

All ML practitioners interviewed for this study had experience outsourcing data-related tasks with both crowdsourcing platforms and BPOs. They all agreed that platforms are cheaper than BPOs, but the latter offer higher quality. As argued by our interview partners, BPO teams remain more or less unchanged throughout the production project, which results in better quality. Moreover, direct communication with project managers allows for iterations and the incorporation of feedback. Several ML practitioners also report preferring not to outsource data-related tasks, especially in cases where a unique “feel for the data” \cite{passi2017}, that can only be achieved with time and experience, was required. The evidence pointing to a negative correlation between cheap labor and the quality of data \cite{litman2015} described by the ML practitioners that we interviewed could be a strong argument for requesters to take measures and fight precarious work in outsourced facilities. Improving labor conditions might result in less expensive (and, perhaps, more effective) approach than investing in “debiasing” datasets after production. 

\subsubsection{Fighting Workers' Surveillance and Control by Encouraging Interrogation}
\hspace{1cm}

Our findings show that the widespread use of “protected categories”  for human classification is bound to the cultural contexts and local jurisdictions that define what counts as a protected group. Moreover, even tasks that do not involve classifying humans, such as identifying objects in a road, can potentially have fatal consequences for individuals or groups, as in the case of the Tasksource requester who did not include labels for humans sleeping or lying on the streets. Making the rationale behind task instructions explicit can be difficult if categories are implicitly considered commonplace for requesters, as they might not even notice the normativity behind instructed taxonomies. Moreover, data workers that are subject to surveillance and control and who risk being banned from tasks are less likely to question instructions. De-centering the development of taxonomies from an “a priori” (i.e., classifying exclusively based on personal experience) and data-based (i.e., classifying solely based on quantitative data) classification to one that derives from the context and experiences of those who may be affected by it could be a fruitful approach to this issue \cite{dignazio2020}. Data workers often perceive errors in task instructions or interfaces that remain unnoticed by the requesters. Even if this feedback could be valuable for requesters, the data-production dispositif is designed to silence workers’ voices.  

We argue that the approach observed in Clickrating, where feedback from workers was encouraged, could be constructive here. However, expanding and implementing such an approach would require a general shift of perspective: from considering workers’ subjectivities a danger to data towards considering workers as assets in the quest for producing high-quality datasets.  Fostering workers' agency instead of surveillance and opening up channels for feedback could allow workers to become co-producers of datasets instead of mere reproduction tools.

\emph{Why would requesters want to be questioned in their logic?}

While taxonomies respond to the commercial necessities of requesters, they also need to be built with equity and inclusivity in mind. This is not only an ethical issue, but it can quickly become a commercial one. Public scrutiny can have fatal consequences for a machine learning product that is perceived to be discriminatory or harmful \cite{Angwin2016,Kim2019,Hao2020a}. Furthermore, instruction documents are living documents. We have observed how requesters update them by withdrawing the tasks, reinstating the instructions, and seeking data work again, a time-consuming and costly process. Thus, requesters could benefit from considering instructions as the product of exchanges with the different stakeholders contributing to data production and deployment. Data workers could play a key role in interrogating and improving tasks and, therefore, datasets and ML systems.

\subsection{Limitations and Future Research}

Our findings are bound to the platforms, companies, individuals, and geographical contexts covered by our study and our positionality as researchers, which has undoubtedly oriented but probably also limited our observations and interpretations. Because of the qualitative nature of our investigation, we have striven for inter-subject comprehensibility \cite{flick2007} instead of objectivity, which means making sure that our interpretations are plausible for both authors and the contexts observed. Furthermore, the use of multiple data sources allowed us to procure supporting evidence for observed phenomena. In addition, the use of expert interviews allowed us to confirm and discuss several of our initial interpretations.

This paper only covers some aspects of the data-production dispositif. This is because no dispositif works in isolation but is always entangled with other discourse, action, and materialization networks. To explicate the totality of the data-production dispositif would mean to analyze its relationship with, among many others, the scientific dispositif, the economic dispositif, and more specifically, the academic and the tech-industry dispositifs. Critical aspects of these relationships have been reported in these pages, but covering them all in one paper would be unfeasible.  The fact that our analysis is bound to remain “incomplete” could be seen as a limitation.  However, we consider it an opportunity for future research to expand our findings and interrogate ways of working with data that today seem commonplace. We think that a profound exploration into the tech-industry dispositif and its relationship with the data-production dispositif could be especially fruitful.

\section{Conclusion}

To explore how data for machine learning is produced through labor outsourced to Venezuela and Argentina, we have turned to the Foucauldian notion of dispositif and applied an adapted version of the dispositif analysis method outlined, among others, by Sigfried Jäger \cite{jaeger2007, jager2016analysing}. Our investigation comprised the analysis of task instructions, interviews with data workers, managers, and requesters, as well as observations at crowdsourcing platforms and a business process outsourcing company.  What we have called the data-production dispositif comprises discourses, work practices, and materializations that are (re)produced in and through ML data work. Our findings have shown that requesters use task instructions to impose predefined forms of interpreting data. The context of poverty and dependence in Latin America leaves workers with no other option but to obey.  
In view of these findings, we propose three ways of counteracting the data-production dispositif and its effects: making worldviews encoded in task instructions explicit, thinking of workers as assets, and empowering them to produce better data. 

While the potentially harmful effects of algorithmic biases continue to be widely discussed, it is also essential to address how power imbalances and imposed classification principles in data creation contribute to the (re)production of inequalities by machine learning. The empowerment of workers and the decommodification of their labor away from market dependency, as well as the detailed documentation of outsourced processes of data creation, remain essential steps to allow spaces of reflection, deliberation, and audit that could potentially contribute to addressing some of the social questions surrounding machine learning technologies.

\begin{acks}
Funded by the German Federal Ministry of Education and Research (BMBF) – Nr 16DII113, the International Development Research Centre of Canada, and the Schwartz Reisman Institute for Technology and Society. We thank Tianling Yang, Marc Pohl, Alex Taylor, Alessandro Delfanti, Paula Núñez de Villavicencio, Paola Tubaro, Antonio Casilli, and the anonymous reviewers.  Special thanks to the data workers who shared their experiences with us. This work would not have been possible without them.
\end{acks}

\bibliographystyle{ACM-Reference-Format}
\bibliography{references}


\begin{thebibliography}{90}


\ifx \showCODEN    \undefined \def \showCODEN     #1{\unskip}     \fi
\ifx \showDOI      \undefined \def \showDOI       #1{#1}\fi
\ifx \showISBNx    \undefined \def \showISBNx     #1{\unskip}     \fi
\ifx \showISBNxiii \undefined \def \showISBNxiii  #1{\unskip}     \fi
\ifx \showISSN     \undefined \def \showISSN      #1{\unskip}     \fi
\ifx \showLCCN     \undefined \def \showLCCN      #1{\unskip}     \fi
\ifx \shownote     \undefined \def \shownote      #1{#1}          \fi
\ifx \showarticletitle \undefined \def \showarticletitle #1{#1}   \fi
\ifx \showURL      \undefined \def \showURL       {\relax}        \fi
\providecommand\bibfield[2]{#2}
\providecommand\bibinfo[2]{#2}
\providecommand\natexlab[1]{#1}
\providecommand\showeprint[2][]{arXiv:#2}

\bibitem[\protect\citeauthoryear{{Agence France-Presse}}{{Agence
  France-Presse}}{2021}]%
        {AgenceFrance-Presse2021}
\bibfield{author}{\bibinfo{person}{{Agence France-Presse}}.}
  \bibinfo{year}{2021}\natexlab{}.
\newblock \showarticletitle{{Venezuela reports 2020 inflation of 3,000
  percent}}.
\newblock \bibinfo{journal}{\emph{ABS CBN News}} (\bibinfo{year}{2021}).
\newblock


\bibitem[\protect\citeauthoryear{Angwin, Larson, Mattu, and Kirchner}{Angwin
  et~al\mbox{.}}{2016}]%
        {Angwin2016}
\bibfield{author}{\bibinfo{person}{Julia Angwin}, \bibinfo{person}{Jeff
  Larson}, \bibinfo{person}{Surya Mattu}, {and} \bibinfo{person}{Lauren
  Kirchner}.} \bibinfo{year}{2016}\natexlab{}.
\newblock \showarticletitle{{Machine Bias}}.
\newblock \bibinfo{journal}{\emph{ProPublica}} (\bibinfo{year}{2016}).
\newblock


\bibitem[\protect\citeauthoryear{Artstein and Poesio}{Artstein and
  Poesio}{2005}]%
        {artstein2005}
\bibfield{author}{\bibinfo{person}{Ron Artstein} {and} \bibinfo{person}{Massimo
  Poesio}.} \bibinfo{year}{2005}\natexlab{}.
\newblock \showarticletitle{Bias decreases in proportion to the number of
  annotators}. In \bibinfo{booktitle}{\emph{Proceedings of {FG}-{MoL} 2005 :
  the 10th {Conference} on {Formal} {Grammar} and the 9th {Meeting} on
  {Mathematics} of {Language}, {Edinburgh}, 5–7 {August}, 2005}}.
  \bibinfo{pages}{139--148}.
\newblock


\bibitem[\protect\citeauthoryear{Bardzell, Bardzell, Zhang, and Pace}{Bardzell
  et~al\mbox{.}}{2014}]%
        {bardzell2014}
\bibfield{author}{\bibinfo{person}{Jeffrey Bardzell}, \bibinfo{person}{Shaowen
  Bardzell}, \bibinfo{person}{Guo Zhang}, {and} \bibinfo{person}{Tyler Pace}.}
  \bibinfo{year}{2014}\natexlab{}.
\newblock \showarticletitle{The lonely raccoon at the ball: designing for
  intimacy, sociability, and selfhood}. In
  \bibinfo{booktitle}{\emph{Proceedings of the {SIGCHI} {Conference} on {Human}
  {Factors} in {Computing} {Systems}}}. \bibinfo{publisher}{ACM},
  \bibinfo{address}{Toronto Ontario Canada}, \bibinfo{pages}{3943--3952}.
\newblock
\showISBNx{978-1-4503-2473-1}
\urldef\tempurl%
\url{https://doi.org/10.1145/2556288.2557127}
\showDOI{\tempurl}


\bibitem[\protect\citeauthoryear{Bender, Gebru, McMillan-Major, and
  Shmitchell}{Bender et~al\mbox{.}}{2021}]%
        {Bender2021}
\bibfield{author}{\bibinfo{person}{Emily~M. Bender}, \bibinfo{person}{Timnit
  Gebru}, \bibinfo{person}{Angelina McMillan-Major}, {and}
  \bibinfo{person}{Shmargaret Shmitchell}.} \bibinfo{year}{2021}\natexlab{}.
\newblock \showarticletitle{{On the Dangers of Stochastic Parrots: Can Language
  Models Be Too Big?}}
\newblock \bibinfo{journal}{\emph{Conference on Fairness, Accountability, and
  Transparency (FAccT '21)}} (\bibinfo{date}{mar} \bibinfo{year}{2021}).
\newblock


\bibitem[\protect\citeauthoryear{Bossen, Pine, Cabitza, Ellingsen, and
  Piras}{Bossen et~al\mbox{.}}{2019}]%
        {bossen2019}
\bibfield{author}{\bibinfo{person}{Claus Bossen}, \bibinfo{person}{Kathleen~H
  Pine}, \bibinfo{person}{Federico Cabitza}, \bibinfo{person}{Gunnar
  Ellingsen}, {and} \bibinfo{person}{Enrico~Maria Piras}.}
  \bibinfo{year}{2019}\natexlab{}.
\newblock \showarticletitle{Data work in healthcare: {An} {Introduction}}.
\newblock \bibinfo{journal}{\emph{Health Informatics Journal}}
  \bibinfo{volume}{25}, \bibinfo{number}{3} (\bibinfo{date}{Sept.}
  \bibinfo{year}{2019}), \bibinfo{pages}{465--474}.
\newblock
\showISSN{1460-4582, 1741-2811}
\urldef\tempurl%
\url{https://doi.org/10.1177/1460458219864730}
\showDOI{\tempurl}


\bibitem[\protect\citeauthoryear{Brodley and Friedl}{Brodley and
  Friedl}{1999}]%
        {brodley1999}
\bibfield{author}{\bibinfo{person}{C.~E. Brodley} {and} \bibinfo{person}{M.~A.
  Friedl}.} \bibinfo{year}{1999}\natexlab{}.
\newblock \showarticletitle{Identifying {Mislabeled} {Training} {Data}}.
\newblock \bibinfo{journal}{\emph{Journal of Artificial Intelligence Research}}
   \bibinfo{volume}{11} (\bibinfo{date}{Aug.} \bibinfo{year}{1999}),
  \bibinfo{pages}{131--167}.
\newblock
\showISSN{1076-9757}
\urldef\tempurl%
\url{https://doi.org/10.1613/jair.606}
\showDOI{\tempurl}


\bibitem[\protect\citeauthoryear{Bührmann and Schneider}{Bührmann and
  Schneider}{2007}]%
        {buehrmann2007}
\bibfield{author}{\bibinfo{person}{Andrea~D. Bührmann} {and}
  \bibinfo{person}{Werner Schneider}.} \bibinfo{year}{2007}\natexlab{}.
\newblock \showarticletitle{Mehr als nur diskursive {Praxis}? –
  {Konzeptionelle} {Grundlagen} und methodische {Aspekte} der
  {Dispositivanalyse}}.
\newblock \bibinfo{journal}{\emph{Forum Qualitative Sozialforschung / Forum:
  Qualitative Social Research}} \bibinfo{volume}{Vol 8}, \bibinfo{number}{No 2:
  From Michel Foucault's Theory of Discourse to Empirical Discourse Research}
  (\bibinfo{date}{May} \bibinfo{year}{2007}).
\newblock
\urldef\tempurl%
\url{https://doi.org/10.17169/FQS-8.2.237}
\showDOI{\tempurl}


\bibitem[\protect\citeauthoryear{Caborn}{Caborn}{2016}]%
        {caborn2016}
\bibfield{author}{\bibinfo{person}{Joannah Caborn}.}
  \bibinfo{year}{2016}\natexlab{}.
\newblock \showarticletitle{On the {Methodology} of {Dispositive} {Analysis}}.
\newblock \bibinfo{journal}{\emph{Critical Approaches to Discourse Analysis
  Across Disciplines}} \bibinfo{volume}{1}, \bibinfo{number}{1}
  (\bibinfo{year}{2016}), \bibinfo{pages}{115--123}.
\newblock
\showISSN{1576-4737}
\urldef\tempurl%
\url{https://doi.org/10.5209/CLAC.53494}
\showDOI{\tempurl}


\bibitem[\protect\citeauthoryear{Casilli}{Casilli}{2017}]%
        {Casilli2017a}
\bibfield{author}{\bibinfo{person}{Antonio~A. Casilli}.}
  \bibinfo{year}{2017}\natexlab{}.
\newblock \showarticletitle{{Digital labor studies go global: Toward a digital
  decolonial turn}}.
\newblock \bibinfo{journal}{\emph{International Journal of Communication}}
  \bibinfo{volume}{11} (\bibinfo{year}{2017}), \bibinfo{pages}{3934--3954}.
\newblock
\showISSN{19328036}


\bibitem[\protect\citeauthoryear{Casilli and Posada}{Casilli and
  Posada}{2019}]%
        {Casilli2019}
\bibfield{author}{\bibinfo{person}{Antonio~A. Casilli} {and}
  \bibinfo{person}{Julian Posada}.} \bibinfo{year}{2019}\natexlab{}.
\newblock \showarticletitle{{The Platformisation of Labor and Society}}.
\newblock In \bibinfo{booktitle}{\emph{Society and the Internet}
  (\bibinfo{edition}{vol. 2} ed.)}, \bibfield{editor}{\bibinfo{person}{Mark
  Graham} {and} \bibinfo{person}{William~H. Dutton}} (Eds.).
  \bibinfo{publisher}{Oxford University Press}, \bibinfo{address}{Oxford}.
\newblock


\bibitem[\protect\citeauthoryear{Cheng and Cosley}{Cheng and Cosley}{2013}]%
        {cheng2013}
\bibfield{author}{\bibinfo{person}{Justin Cheng} {and} \bibinfo{person}{Dan
  Cosley}.} \bibinfo{year}{2013}\natexlab{}.
\newblock \showarticletitle{How annotation styles influence content and
  preferences}. In \bibinfo{booktitle}{\emph{Proceedings of the 24th {ACM}
  {Conference} on {Hypertext} and {Social} {Media} - {HT} '13}}.
  \bibinfo{publisher}{Association for Computing Machinery},
  \bibinfo{address}{Paris, France}, \bibinfo{pages}{214--218}.
\newblock
\showISBNx{978-1-4503-1967-6}
\urldef\tempurl%
\url{https://doi.org/10.1145/2481492.2481519}
\showDOI{\tempurl}
\newblock
\shownote{tex.ids: cheng2013a.}


\bibitem[\protect\citeauthoryear{Crawford}{Crawford}{2021}]%
        {Crawford2021}
\bibfield{author}{\bibinfo{person}{Kate Crawford}.}
  \bibinfo{year}{2021}\natexlab{}.
\newblock \bibinfo{booktitle}{\emph{{Atlas of AI. Power, Politics, and the
  Planetary Costs of Artificial Intelligence}}}.
\newblock \bibinfo{publisher}{Yale University Press}, \bibinfo{address}{New
  Haven, CT}.
\newblock


\bibitem[\protect\citeauthoryear{Cronin}{Cronin}{1996}]%
        {cronin1996}
\bibfield{author}{\bibinfo{person}{Ciaran Cronin}.}
  \bibinfo{year}{1996}\natexlab{}.
\newblock \showarticletitle{Bourdieu and {Foucault} on power and modernity}.
\newblock \bibinfo{journal}{\emph{Philosophy \& Social Criticism}}
  \bibinfo{volume}{22}, \bibinfo{number}{6} (\bibinfo{date}{Nov.}
  \bibinfo{year}{1996}), \bibinfo{pages}{55--85}.
\newblock
\showISSN{0191-4537, 1461-734X}
\urldef\tempurl%
\url{https://doi.org/10.1177/019145379602200603}
\showDOI{\tempurl}


\bibitem[\protect\citeauthoryear{Dauvergne}{Dauvergne}{2020}]%
        {Dauvergne2020}
\bibfield{author}{\bibinfo{person}{Peter Dauvergne}.}
  \bibinfo{year}{2020}\natexlab{}.
\newblock \bibinfo{booktitle}{\emph{{AI in the Wild. Sustainability in the Age
  of Artificial Intelligence}}}.
\newblock \bibinfo{publisher}{MIT Press}, \bibinfo{address}{Cambridge, MA}.
\newblock


\bibitem[\protect\citeauthoryear{Delfanti and Sharma}{Delfanti and
  Sharma}{2019}]%
        {Delfanti2019a}
\bibfield{editor}{\bibinfo{person}{Alessandro Delfanti} {and}
  \bibinfo{person}{Sarah Sharma}} (Eds.). \bibinfo{year}{2019}\natexlab{}.
\newblock \showarticletitle{{Log Out! The Platform Economy and Worker
  Resistance}}.
\newblock \bibinfo{journal}{\emph{Notes from Below}} \bibinfo{number}{8}
  (\bibinfo{year}{2019}).
\newblock


\bibitem[\protect\citeauthoryear{D'Ignazio and Klein}{D'Ignazio and
  Klein}{2020}]%
        {dignazio2020}
\bibfield{author}{\bibinfo{person}{Catherine D'Ignazio} {and}
  \bibinfo{person}{Lauren~F. Klein}.} \bibinfo{year}{2020}\natexlab{}.
\newblock \bibinfo{booktitle}{\emph{Data feminism}}.
\newblock \bibinfo{publisher}{The MIT Press}, \bibinfo{address}{Cambridge,
  Massachusetts}.
\newblock
\showISBNx{978-0-262-04400-4}
\urldef\tempurl%
\url{https://mitpress.mit.edu/books/data-feminism}
\showURL{%
\tempurl}


\bibitem[\protect\citeauthoryear{Fan, Gadiraju, Checco, and Demartini}{Fan
  et~al\mbox{.}}{2020}]%
        {fan2020}
\bibfield{author}{\bibinfo{person}{Shaoyang Fan}, \bibinfo{person}{Ujwal
  Gadiraju}, \bibinfo{person}{Alessandro Checco}, {and}
  \bibinfo{person}{Gianluca Demartini}.} \bibinfo{year}{2020}\natexlab{}.
\newblock \showarticletitle{{CrowdCO}-{OP}: {Sharing} {Risks} and {Rewards} in
  {Crowdsourcing}}.
\newblock \bibinfo{journal}{\emph{Proceedings of the ACM on Human-Computer
  Interaction}} \bibinfo{volume}{4}, \bibinfo{number}{CSCW2}
  (\bibinfo{date}{Oct.} \bibinfo{year}{2020}), \bibinfo{pages}{1--24}.
\newblock
\showISSN{2573-0142}
\urldef\tempurl%
\url{https://doi.org/10.1145/3415203}
\showDOI{\tempurl}


\bibitem[\protect\citeauthoryear{Feinberg}{Feinberg}{2017}]%
        {feinberg2017}
\bibfield{author}{\bibinfo{person}{Melanie Feinberg}.}
  \bibinfo{year}{2017}\natexlab{}.
\newblock \showarticletitle{A {Design} {Perspective} on {Data}}. In
  \bibinfo{booktitle}{\emph{{CHI} '17: {Proceedings} of the 2017 {CHI}
  {Conference} on {Human} {Factors} in {Computing} {Systems}}}
  \emph{(\bibinfo{series}{{CHI} ’17})}. \bibinfo{publisher}{Association for
  Computing Machinery}, \bibinfo{address}{Denver, Colorado, USA},
  \bibinfo{pages}{2952--2963}.
\newblock
\showISBNx{978-1-4503-4655-9}
\urldef\tempurl%
\url{https://doi.org/10.1145/3025453.3025837}
\showDOI{\tempurl}


\bibitem[\protect\citeauthoryear{Flick}{Flick}{2007}]%
        {flick2007}
\bibfield{author}{\bibinfo{person}{Uwe Flick}.}
  \bibinfo{year}{2007}\natexlab{}.
\newblock \bibinfo{booktitle}{\emph{Qualitative {Sozialforschung}: {Eine}
  {Einführung}} (\bibinfo{edition}{10, erweiterte neuausgabe} ed.)}.
\newblock \bibinfo{publisher}{Rowohlt Taschenbuch}, \bibinfo{address}{Reinbek
  bei Hamburg}.
\newblock
\showISBNx{978-3-499-55694-4}


\bibitem[\protect\citeauthoryear{Foucault}{Foucault}{1971}]%
        {foucault1971}
\bibfield{author}{\bibinfo{person}{Michel Foucault}.}
  \bibinfo{year}{1971}\natexlab{}.
\newblock \showarticletitle{Orders of discourse}.
\newblock \bibinfo{journal}{\emph{Social Science Information}}
  \bibinfo{volume}{10}, \bibinfo{number}{2} (\bibinfo{date}{April}
  \bibinfo{year}{1971}), \bibinfo{pages}{7--30}.
\newblock
\showISSN{0539-0184}
\urldef\tempurl%
\url{https://doi.org/10.1177/053901847101000201}
\showDOI{\tempurl}
\newblock
\shownote{Publisher: SAGE Publications Ltd.}


\bibitem[\protect\citeauthoryear{Foucault}{Foucault}{1982a}]%
        {foucault1982a}
\bibfield{author}{\bibinfo{person}{Michel Foucault}.}
  \bibinfo{year}{1982}\natexlab{a}.
\newblock \bibinfo{booktitle}{\emph{The {Archaeology} of {Knowledge}: {And} the
  {Discourse} on {Language}}}.
\newblock \bibinfo{publisher}{Vintage}, \bibinfo{address}{New York}.
\newblock
\showISBNx{978-0-394-71106-5}


\bibitem[\protect\citeauthoryear{Foucault}{Foucault}{1982b}]%
        {foucault1982}
\bibfield{author}{\bibinfo{person}{Michel Foucault}.}
  \bibinfo{year}{1982}\natexlab{b}.
\newblock \showarticletitle{The {Subject} and {Power}}.
\newblock \bibinfo{journal}{\emph{Critical Inquiry}} \bibinfo{volume}{8},
  \bibinfo{number}{4} (\bibinfo{year}{1982}), \bibinfo{pages}{777--795}.
\newblock
\urldef\tempurl%
\url{https://www.jstor.org/stable/1343197}
\showURL{%
\tempurl}


\bibitem[\protect\citeauthoryear{Foucault}{Foucault}{1996}]%
        {Foucault1996-FOUWIC}
\bibfield{author}{\bibinfo{person}{Michel Foucault}.}
  \bibinfo{year}{1996}\natexlab{}.
\newblock \showarticletitle{What Is Critique?}
\newblock In \bibinfo{booktitle}{\emph{What is Enlightenment?:
  Eighteenth-Century Answers and Twentieth-Century Questions}},
  \bibfield{editor}{\bibinfo{person}{James Schmidt}} (Ed.).
  \bibinfo{publisher}{University of California Press}.
\newblock


\bibitem[\protect\citeauthoryear{Foucault and Gordon}{Foucault and
  Gordon}{1980}]%
        {foucault1980}
\bibfield{author}{\bibinfo{person}{Michel Foucault} {and}
  \bibinfo{person}{Colin Gordon}.} \bibinfo{year}{1980}\natexlab{}.
\newblock \bibinfo{booktitle}{\emph{Power/knowledge: selected interviews and
  other writings, 1972-1977} (\bibinfo{edition}{1st american ed} ed.)}.
\newblock \bibinfo{publisher}{Pantheon Books}, \bibinfo{address}{New York}.
\newblock
\showISBNx{978-0-394-51357-7 978-0-394-73954-0}


\bibitem[\protect\citeauthoryear{Gadiraju, Yang, and Bozzon}{Gadiraju
  et~al\mbox{.}}{2017}]%
        {gadiraju2017}
\bibfield{author}{\bibinfo{person}{Ujwal Gadiraju}, \bibinfo{person}{Jie Yang},
  {and} \bibinfo{person}{Alessandro Bozzon}.} \bibinfo{year}{2017}\natexlab{}.
\newblock \showarticletitle{Clarity is a {Worthwhile} {Quality}: {On} the
  {Role} of {Task} {Clarity} in {Microtask} {Crowdsourcing}}. In
  \bibinfo{booktitle}{\emph{Proceedings of the 28th {ACM} {Conference} on
  {Hypertext} and {Social} {Media}}} \emph{(\bibinfo{series}{{HT} '17})}.
  \bibinfo{publisher}{Association for Computing Machinery},
  \bibinfo{address}{New York, NY, USA}, \bibinfo{pages}{5--14}.
\newblock
\showISBNx{978-1-4503-4708-2}
\urldef\tempurl%
\url{https://doi.org/10.1145/3078714.3078715}
\showDOI{\tempurl}


\bibitem[\protect\citeauthoryear{Geiger, Yu, Yang, Dai, Qiu, Tang, and
  Huang}{Geiger et~al\mbox{.}}{2020}]%
        {geiger2020}
\bibfield{author}{\bibinfo{person}{R.~Stuart Geiger}, \bibinfo{person}{Kevin
  Yu}, \bibinfo{person}{Yanlai Yang}, \bibinfo{person}{Mindy Dai},
  \bibinfo{person}{Jie Qiu}, \bibinfo{person}{Rebekah Tang}, {and}
  \bibinfo{person}{Jenny Huang}.} \bibinfo{year}{2020}\natexlab{}.
\newblock \showarticletitle{Garbage in, Garbage out? Do Machine Learning
  Application Papers in Social Computing Report Where Human-Labeled Training
  Data Comes From?}. In \bibinfo{booktitle}{\emph{Proceedings of the 2020
  Conference on Fairness, Accountability, and Transparency}} (Barcelona, Spain)
  \emph{(\bibinfo{series}{FAT* '20})}. \bibinfo{publisher}{Association for
  Computing Machinery}, \bibinfo{address}{New York, NY, USA},
  \bibinfo{pages}{325–336}.
\newblock
\showISBNx{9781450369367}
\urldef\tempurl%
\url{https://doi.org/10.1145/3351095.3372862}
\showDOI{\tempurl}


\bibitem[\protect\citeauthoryear{Geva, Goldberg, and Berant}{Geva
  et~al\mbox{.}}{2019}]%
        {geva2019}
\bibfield{author}{\bibinfo{person}{Mor Geva}, \bibinfo{person}{Yoav Goldberg},
  {and} \bibinfo{person}{Jonathan Berant}.} \bibinfo{year}{2019}\natexlab{}.
\newblock \showarticletitle{Are {We} {Modeling} the {Task} or the {Annotator}?
  {An} {Investigation} of {Annotator} {Bias} in {Natural} {Language}
  {Understanding} {Datasets}}. In \bibinfo{booktitle}{\emph{Proceedings of the
  2019 {Conference} on {Empirical} {Methods} in {Natural} {Language}
  {Processing} and the 9th {International} {Joint} {Conference} on {Natural}
  {Language} {Processing} ({EMNLP}-{IJCNLP})}}. \bibinfo{publisher}{Association
  for Computational Linguistics}, \bibinfo{address}{Hong Kong, China},
  \bibinfo{pages}{1161--1166}.
\newblock
\urldef\tempurl%
\url{https://doi.org/10.18653/v1/D19-1107}
\showDOI{\tempurl}


\bibitem[\protect\citeauthoryear{Ghai, Liao, Zhang, and Mueller}{Ghai
  et~al\mbox{.}}{2020}]%
        {ghai2020}
\bibfield{author}{\bibinfo{person}{Bhavya Ghai}, \bibinfo{person}{Q.~Vera
  Liao}, \bibinfo{person}{Yunfeng Zhang}, {and} \bibinfo{person}{Klaus
  Mueller}.} \bibinfo{year}{2020}\natexlab{}.
\newblock \showarticletitle{Measuring {Social} {Biases} of {Crowd} {Workers}
  using {Counterfactual} {Queries}}. \bibinfo{address}{Honolulu, HI, USA}.
\newblock
\urldef\tempurl%
\url{http://fair-ai.owlstown.com/publications/1424}
\showURL{%
\tempurl}


\bibitem[\protect\citeauthoryear{Gray and Suri}{Gray and Suri}{2019}]%
        {Gray2019}
\bibfield{author}{\bibinfo{person}{Mary~L. Gray} {and}
  \bibinfo{person}{Siddharth Suri}.} \bibinfo{year}{2019}\natexlab{}.
\newblock \bibinfo{booktitle}{\emph{Ghost {Work}: {How} to {Stop} {Silicon}
  {Valley} from {Building} a {New} {Global} {Underclass}}}.
\newblock \bibinfo{publisher}{Houghton Mifflin Harcourt},
  \bibinfo{address}{Boston}.
\newblock
\showISBNx{978-1-328-56624-9}


\bibitem[\protect\citeauthoryear{Hamann, Maesse, Scholz, and
  Angermuller}{Hamann et~al\mbox{.}}{2019}]%
        {hamann2019}
\bibfield{author}{\bibinfo{person}{Julian Hamann}, \bibinfo{person}{Jens
  Maesse}, \bibinfo{person}{Ronny Scholz}, {and} \bibinfo{person}{Johannes
  Angermuller}.} \bibinfo{year}{2019}\natexlab{}.
\newblock \showarticletitle{The {Academic} {Dispositif}: {Towards} a
  {Context}-{Centred} {Discourse} {Analysis}}.
\newblock In \bibinfo{booktitle}{\emph{Quantifying {Approaches} to {Discourse}
  for {Social} {Scientists}}}, \bibfield{editor}{\bibinfo{person}{Ronny
  Scholz}} (Ed.). \bibinfo{publisher}{Springer International Publishing},
  \bibinfo{address}{Cham}, \bibinfo{pages}{51--87}.
\newblock
\showISBNx{978-3-319-97369-2 978-3-319-97370-8}
\urldef\tempurl%
\url{https://doi.org/10.1007/978-3-319-97370-8_3}
\showDOI{\tempurl}


\bibitem[\protect\citeauthoryear{Hao}{Hao}{2020}]%
        {Hao2020a}
\bibfield{author}{\bibinfo{person}{Karen Hao}.}
  \bibinfo{year}{2020}\natexlab{}.
\newblock \showarticletitle{{In 2020, let's stop AI ethics-washing and actually
  do something}}.
\newblock \bibinfo{journal}{\emph{MIT Technology Review}}
  (\bibinfo{year}{2020}).
\newblock
\urldef\tempurl%
\url{https://www.technologyreview.com/2019/12/27/57/ai-ethics-washing-time-to-act/}
\showURL{%
\tempurl}


\bibitem[\protect\citeauthoryear{Hara, Adams, Milland, Savage, Callison-Burch,
  and Bigham}{Hara et~al\mbox{.}}{2018}]%
        {hara2018}
\bibfield{author}{\bibinfo{person}{Kotaro Hara}, \bibinfo{person}{Abigail
  Adams}, \bibinfo{person}{Kristy Milland}, \bibinfo{person}{Saiph Savage},
  \bibinfo{person}{Chris Callison-Burch}, {and} \bibinfo{person}{Jeffrey~P.
  Bigham}.} \bibinfo{year}{2018}\natexlab{}.
\newblock \showarticletitle{A {Data}-{Driven} {Analysis} of {Workers}'
  {Earnings} on {Amazon} {Mechanical} {Turk}}. In
  \bibinfo{booktitle}{\emph{Proceedings of the 2018 {CHI} {Conference} on
  {Human} {Factors} in {Computing} {Systems}}}. \bibinfo{publisher}{ACM},
  \bibinfo{address}{Montreal QC Canada}, \bibinfo{pages}{1--14}.
\newblock
\showISBNx{978-1-4503-5620-6}
\urldef\tempurl%
\url{https://doi.org/10.1145/3173574.3174023}
\showDOI{\tempurl}


\bibitem[\protect\citeauthoryear{Harmon and Mazmanian}{Harmon and
  Mazmanian}{2013}]%
        {harmon2013}
\bibfield{author}{\bibinfo{person}{Ellie Harmon} {and} \bibinfo{person}{Melissa
  Mazmanian}.} \bibinfo{year}{2013}\natexlab{}.
\newblock \showarticletitle{Stories of the {Smartphone} in everyday discourse:
  conflict, tension \& instability}. In \bibinfo{booktitle}{\emph{Proceedings
  of the {SIGCHI} {Conference} on {Human} {Factors} in {Computing} {Systems}}}.
  \bibinfo{publisher}{ACM}, \bibinfo{address}{Paris France},
  \bibinfo{pages}{1051--1060}.
\newblock
\showISBNx{978-1-4503-1899-0}
\urldef\tempurl%
\url{https://doi.org/10.1145/2470654.2466134}
\showDOI{\tempurl}


\bibitem[\protect\citeauthoryear{Horst and Miller}{Horst and Miller}{2012}]%
        {Horst2012a}
\bibfield{author}{\bibinfo{person}{Heather~A. Horst} {and}
  \bibinfo{person}{Daniel Miller}.} \bibinfo{year}{2012}\natexlab{}.
\newblock \bibinfo{booktitle}{\emph{{Digital Anthropology}}}.
\newblock \bibinfo{publisher}{Berg}. 196--213 pages.
\newblock


\bibitem[\protect\citeauthoryear{Hube, Fetahu, and Gadiraju}{Hube
  et~al\mbox{.}}{2019}]%
        {hube2019}
\bibfield{author}{\bibinfo{person}{Christoph Hube}, \bibinfo{person}{Besnik
  Fetahu}, {and} \bibinfo{person}{Ujwal Gadiraju}.}
  \bibinfo{year}{2019}\natexlab{}.
\newblock \showarticletitle{Understanding and {Mitigating} {Worker} {Biases} in
  the {Crowdsourced} {Collection} of {Subjective} {Judgments}}. In
  \bibinfo{booktitle}{\emph{Proceedings of the 2019 {CHI} {Conference} on
  {Human} {Factors} in {Computing} {Systems}}} \emph{(\bibinfo{series}{{CHI}
  '19})}. \bibinfo{publisher}{Association for Computing Machinery},
  \bibinfo{address}{New York, NY, USA}, \bibinfo{pages}{1--12}.
\newblock
\showISBNx{978-1-4503-5970-2}
\urldef\tempurl%
\url{https://doi.org/10.1145/3290605.3300637}
\showDOI{\tempurl}
\newblock
\shownote{tex.ids: hube2019a event-place: Glasgow, Scotland Uk.}


\bibitem[\protect\citeauthoryear{Irani}{Irani}{2015}]%
        {Irani2015}
\bibfield{author}{\bibinfo{person}{Lilly Irani}.}
  \bibinfo{year}{2015}\natexlab{}.
\newblock \showarticletitle{{The cultural work of microwork}}.
\newblock \bibinfo{journal}{\emph{New Media {\&} Society}}
  \bibinfo{volume}{17}, \bibinfo{number}{5} (\bibinfo{year}{2015}),
  \bibinfo{pages}{720--739}.
\newblock
\showISBNx{9780415611152}
\showISSN{1461-4448, 1461-7315}
\urldef\tempurl%
\url{https://doi.org/10.1177/1461444813511926}
\showDOI{\tempurl}


\bibitem[\protect\citeauthoryear{Irani and Silberman}{Irani and
  Silberman}{2013}]%
        {irani2013}
\bibfield{author}{\bibinfo{person}{Lilly~C. Irani} {and}
  \bibinfo{person}{M.~Six Silberman}.} \bibinfo{year}{2013}\natexlab{}.
\newblock \showarticletitle{Turkopticon: interrupting worker invisibility in
  amazon mechanical turk}. In \bibinfo{booktitle}{\emph{Proceedings of the
  {SIGCHI} {Conference} on {Human} {Factors} in {Computing} {Systems}}}
  \emph{(\bibinfo{series}{{CHI} '13})}. \bibinfo{publisher}{Association for
  Computing Machinery}, \bibinfo{address}{Paris, France},
  \bibinfo{pages}{611--620}.
\newblock
\showISBNx{978-1-4503-1899-0}
\urldef\tempurl%
\url{https://doi.org/10.1145/2470654.2470742}
\showDOI{\tempurl}


\bibitem[\protect\citeauthoryear{J{\"a}ger and Maier}{J{\"a}ger and
  Maier}{2016}]%
        {jager2016analysing}
\bibfield{author}{\bibinfo{person}{Siegfried J{\"a}ger} {and}
  \bibinfo{person}{Florentine Maier}.} \bibinfo{year}{2016}\natexlab{}.
\newblock \showarticletitle{Analysing discourses and dispositives: A
  Foucauldian approach to theory and methodology}.
\newblock \bibinfo{journal}{\emph{Methods of critical discourse studies}}
  (\bibinfo{year}{2016}), \bibinfo{pages}{109--136}.
\newblock


\bibitem[\protect\citeauthoryear{Justie}{Justie}{2021}]%
        {Justie2021}
\bibfield{author}{\bibinfo{person}{Brian Justie}.}
  \bibinfo{year}{2021}\natexlab{}.
\newblock \showarticletitle{{Little history of CAPTCHA}}.
\newblock \bibinfo{journal}{\emph{Internet Histories}} \bibinfo{volume}{5},
  \bibinfo{number}{1} (\bibinfo{year}{2021}), \bibinfo{pages}{30--47}.
\newblock
\showISSN{2470-1475}
\urldef\tempurl%
\url{https://doi.org/10.1080/24701475.2020.1831197}
\showDOI{\tempurl}


\bibitem[\protect\citeauthoryear{Jäger}{Jäger}{2007}]%
        {jaeger2007}
\bibfield{author}{\bibinfo{person}{Siegfried Jäger}.}
  \bibinfo{year}{2007}\natexlab{}.
\newblock \bibinfo{booktitle}{\emph{Deutungskämpfe: {Theorie} und {Praxis}
  {Kritischer} {Diskursanalyse}}}.
\newblock \bibinfo{publisher}{Springer-Verlag}.
\newblock
\showISBNx{978-3-531-15072-7}


\bibitem[\protect\citeauthoryear{Kannabiran, Bardzell, and Bardzell}{Kannabiran
  et~al\mbox{.}}{2011}]%
        {kannabiran2011}
\bibfield{author}{\bibinfo{person}{Gopinaath Kannabiran},
  \bibinfo{person}{Jeffrey Bardzell}, {and} \bibinfo{person}{Shaowen
  Bardzell}.} \bibinfo{year}{2011}\natexlab{}.
\newblock \showarticletitle{How {HCI} talks about sexuality: discursive
  strategies, blind spots, and opportunities for future research}. In
  \bibinfo{booktitle}{\emph{Proceedings of the {SIGCHI} {Conference} on {Human}
  {Factors} in {Computing} {Systems}}}. \bibinfo{publisher}{ACM},
  \bibinfo{address}{Vancouver BC Canada}, \bibinfo{pages}{695--704}.
\newblock
\showISBNx{978-1-4503-0228-9}
\urldef\tempurl%
\url{https://doi.org/10.1145/1978942.1979043}
\showDOI{\tempurl}


\bibitem[\protect\citeauthoryear{Katz and Krueger}{Katz and Krueger}{2016}]%
        {Katz2016}
\bibfield{author}{\bibinfo{person}{Lawrence~F. Katz} {and}
  \bibinfo{person}{Alan~B. Krueger}.} \bibinfo{year}{2016}\natexlab{}.
\newblock \bibinfo{title}{{The Role of Unemployment in the Rise in Alternative
  Work Arrangements}}.
\newblock , \bibinfo{numpages}{10}~pages.
\newblock
\showISSN{00028282}
\urldef\tempurl%
\url{https://doi.org/10.1257/aer.p20171092}
\showDOI{\tempurl}


\bibitem[\protect\citeauthoryear{Kazimzade and Miceli}{Kazimzade and
  Miceli}{2020}]%
        {kazimzade_biased_2020}
\bibfield{author}{\bibinfo{person}{Gunay Kazimzade} {and}
  \bibinfo{person}{Milagros Miceli}.} \bibinfo{year}{2020}\natexlab{}.
\newblock \showarticletitle{Biased {Priorities}, {Biased} {Outcomes}: {Three}
  {Recommendations} for {Ethics}-oriented {Data} {Annotation} {Practices}}. In
  \bibinfo{booktitle}{\emph{Proceedings of the {AAAI}/{ACM} {Conference} on
  {Artificial} {Intelligence}, {Ethics}, and {Society}.}}
  \emph{(\bibinfo{series}{{AIES} ’20})}. \bibinfo{publisher}{Association for
  Computing Machinery}, \bibinfo{address}{New York, NY, USA},
  \bibinfo{pages}{1--7}.
\newblock
\showISBNx{978-1-4503-7110-0}
\urldef\tempurl%
\url{https://doi.org/10.1145/3375627.3375809}
\showDOI{\tempurl}


\bibitem[\protect\citeauthoryear{Kim}{Kim}{2019}]%
        {Kim2019}
\bibfield{author}{\bibinfo{person}{Granate Kim}.}
  \bibinfo{year}{2019}\natexlab{}.
\newblock \showarticletitle{{Microsoft Funds Facial Recognition Technology
  Secretly Tested on Palestinians}}.
\newblock \bibinfo{journal}{\emph{Truthout}} (\bibinfo{year}{2019}).
\newblock


\bibitem[\protect\citeauthoryear{Kou, Gui, Chen, and Nardi}{Kou
  et~al\mbox{.}}{2019}]%
        {kou2019}
\bibfield{author}{\bibinfo{person}{Yubo Kou}, \bibinfo{person}{Xinning Gui},
  \bibinfo{person}{Yunan Chen}, {and} \bibinfo{person}{Bonnie Nardi}.}
  \bibinfo{year}{2019}\natexlab{}.
\newblock \showarticletitle{Turn to the {Self} in {Human}-{Computer}
  {Interaction}: {Care} of the {Self} in {Negotiating} the {Human}-{Technology}
  {Relationship}}. In \bibinfo{booktitle}{\emph{Proceedings of the 2019 {CHI}
  {Conference} on {Human} {Factors} in {Computing} {Systems}}}.
  \bibinfo{publisher}{ACM}, \bibinfo{address}{Glasgow Scotland Uk},
  \bibinfo{pages}{1--15}.
\newblock
\showISBNx{978-1-4503-5970-2}
\urldef\tempurl%
\url{https://doi.org/10.1145/3290605.3300711}
\showDOI{\tempurl}


\bibitem[\protect\citeauthoryear{Larroche}{Larroche}{2019}]%
        {larroche2019}
\bibfield{author}{\bibinfo{person}{Valérie Larroche}.}
  \bibinfo{year}{2019}\natexlab{}.
\newblock \bibinfo{booktitle}{\emph{The {Dispositif}: {A} {Concept} for
  {Information} and {Communication} {Sciences}} (\bibinfo{edition}{1} ed.)}.
\newblock \bibinfo{publisher}{Wiley}.
\newblock
\showISBNx{978-1-78630-309-7 978-1-119-50872-4}
\urldef\tempurl%
\url{https://doi.org/10.1002/9781119508724}
\showDOI{\tempurl}


\bibitem[\protect\citeauthoryear{Link}{Link}{2014}]%
        {link2014}
\bibfield{author}{\bibinfo{person}{Jürgen Link}.}
  \bibinfo{year}{2014}\natexlab{}.
\newblock \showarticletitle{Dispositiv}.
\newblock In \bibinfo{booktitle}{\emph{Foucault-{Hanbuch}}},
  \bibfield{editor}{\bibinfo{person}{Clemens Kammler}, \bibinfo{person}{Rolf
  Parr}, \bibinfo{person}{Ulrich~Johannes Schneider}, {and}
  \bibinfo{person}{Elke Reinhardt-Becker}} (Eds.). \bibinfo{publisher}{J.B.
  Metzler}, \bibinfo{address}{Stuttgart}, \bibinfo{pages}{237--242}.
\newblock
\showISBNx{978-3-476-02559-3 978-3-476-01378-1}
\urldef\tempurl%
\url{https://doi.org/10.1007/978-3-476-01378-1_27}
\showDOI{\tempurl}


\bibitem[\protect\citeauthoryear{Litman, Robinson, and Rosenzweig}{Litman
  et~al\mbox{.}}{2015}]%
        {litman2015}
\bibfield{author}{\bibinfo{person}{Leib Litman}, \bibinfo{person}{Jonathan
  Robinson}, {and} \bibinfo{person}{Cheskie Rosenzweig}.}
  \bibinfo{year}{2015}\natexlab{}.
\newblock \showarticletitle{The relationship between motivation, monetary
  compensation, and data quality among {US}- and {India}-based workers on
  {Mechanical} {Turk}}.
\newblock \bibinfo{journal}{\emph{Behavior Research Methods}}
  \bibinfo{volume}{47}, \bibinfo{number}{2} (\bibinfo{date}{June}
  \bibinfo{year}{2015}), \bibinfo{pages}{519--528}.
\newblock
\showISSN{1554-3528}
\urldef\tempurl%
\url{https://doi.org/10.3758/s13428-014-0483-x}
\showDOI{\tempurl}


\bibitem[\protect\citeauthoryear{Lu, Dillahunt, Marcu, and Ackerman}{Lu
  et~al\mbox{.}}{2021}]%
        {Lu2021}
\bibfield{author}{\bibinfo{person}{Alex~Jiahong Lu},
  \bibinfo{person}{Tawanna~R. Dillahunt}, \bibinfo{person}{Gabriela Marcu},
  {and} \bibinfo{person}{Mark~S. Ackerman}.} \bibinfo{year}{2021}\natexlab{}.
\newblock \showarticletitle{{Data Work in Education: Enacting and Negotiating
  Care and Control in Teachers' Use of Data-Driven Classroom Surveillance
  Technology}}.
\newblock \bibinfo{journal}{\emph{Proceedings of the ACM on Human-Computer
  Interaction}} \bibinfo{volume}{5}, \bibinfo{number}{CSCW2}
  (\bibinfo{date}{oct} \bibinfo{year}{2021}), \bibinfo{pages}{1--26}.
\newblock
\showISSN{2573-0142}
\urldef\tempurl%
\url{https://doi.org/10.1145/3479596}
\showDOI{\tempurl}


\bibitem[\protect\citeauthoryear{Manderscheid}{Manderscheid}{2014}]%
        {manderscheid2014}
\bibfield{author}{\bibinfo{person}{Katharina Manderscheid}.}
  \bibinfo{year}{2014}\natexlab{}.
\newblock \showarticletitle{The {Movement} {Problem}, the {Car} and {Future}
  {Mobility} {Regimes}: {Automobility} as {Dispositif} and {Mode} of
  {Regulation}}.
\newblock \bibinfo{journal}{\emph{Mobilities}} \bibinfo{volume}{9},
  \bibinfo{number}{4} (\bibinfo{date}{Oct.} \bibinfo{year}{2014}),
  \bibinfo{pages}{604--626}.
\newblock
\showISSN{1745-0101, 1745-011X}
\urldef\tempurl%
\url{https://doi.org/10.1080/17450101.2014.961257}
\showDOI{\tempurl}


\bibitem[\protect\citeauthoryear{Martin, Hanrahan, O'Neill, and Gupta}{Martin
  et~al\mbox{.}}{2014}]%
        {martin2014}
\bibfield{author}{\bibinfo{person}{David Martin}, \bibinfo{person}{Benjamin~V.
  Hanrahan}, \bibinfo{person}{Jacki O'Neill}, {and} \bibinfo{person}{Neha
  Gupta}.} \bibinfo{year}{2014}\natexlab{}.
\newblock \showarticletitle{Being a turker}. In
  \bibinfo{booktitle}{\emph{Proceedings of the 17th {ACM} conference on
  {Computer} supported cooperative work \& social computing}}.
  \bibinfo{publisher}{ACM}, \bibinfo{address}{Baltimore Maryland USA},
  \bibinfo{pages}{224--235}.
\newblock
\showISBNx{978-1-4503-2540-0}
\urldef\tempurl%
\url{https://doi.org/10.1145/2531602.2531663}
\showDOI{\tempurl}


\bibitem[\protect\citeauthoryear{Miceli, Schuessler, and Yang}{Miceli
  et~al\mbox{.}}{2020}]%
        {miceli2020}
\bibfield{author}{\bibinfo{person}{Milagros Miceli}, \bibinfo{person}{Martin
  Schuessler}, {and} \bibinfo{person}{Tianling Yang}.}
  \bibinfo{year}{2020}\natexlab{}.
\newblock \showarticletitle{Between {Subjectivity} and {Imposition}: {Power}
  {Dynamics} in {Data} {Annotation} for {Computer} {Vision}}.
\newblock \bibinfo{journal}{\emph{Proceedings of the ACM on Human-Computer
  Interaction}} \bibinfo{volume}{4}, \bibinfo{number}{CSCW2}
  (\bibinfo{date}{Oct.} \bibinfo{year}{2020}), \bibinfo{pages}{1--25}.
\newblock
\showISSN{2573-0142, 2573-0142}
\urldef\tempurl%
\url{https://doi.org/10.1145/3415186}
\showDOI{\tempurl}


\bibitem[\protect\citeauthoryear{Miceli, Yang, Naudts, Schuessler, Serbanescu,
  and Hanna}{Miceli et~al\mbox{.}}{2021}]%
        {miceli2021a}
\bibfield{author}{\bibinfo{person}{Milagros Miceli}, \bibinfo{person}{Tianling
  Yang}, \bibinfo{person}{Laurens Naudts}, \bibinfo{person}{Martin Schuessler},
  \bibinfo{person}{Diana Serbanescu}, {and} \bibinfo{person}{Alex Hanna}.}
  \bibinfo{year}{2021}\natexlab{}.
\newblock \showarticletitle{Documenting {Computer} {Vision} {Datasets}: {An}
  {Invitation} to {Reflexive} {Data} {Practices}}. In
  \bibinfo{booktitle}{\emph{Proceedings of the 2021 {ACM} {Conference} on
  {Fairness}, {Accountability}, and {Transparency}}}. \bibinfo{publisher}{ACM},
  \bibinfo{address}{Virtual Event Canada}, \bibinfo{pages}{161--172}.
\newblock
\showISBNx{978-1-4503-8309-7}
\urldef\tempurl%
\url{https://doi.org/10.1145/3442188.3445880}
\showDOI{\tempurl}


\bibitem[\protect\citeauthoryear{M\o{}ller, Bossen, Pine, Nielsen, and
  Neff}{M\o{}ller et~al\mbox{.}}{2020}]%
        {Moller2020}
\bibfield{author}{\bibinfo{person}{Naja~Holten M\o{}ller},
  \bibinfo{person}{Claus Bossen}, \bibinfo{person}{Kathleen~H. Pine},
  \bibinfo{person}{Trine~Rask Nielsen}, {and} \bibinfo{person}{Gina Neff}.}
  \bibinfo{year}{2020}\natexlab{}.
\newblock \showarticletitle{Who Does the Work of Data?}
\newblock \bibinfo{journal}{\emph{Interactions}} \bibinfo{volume}{27},
  \bibinfo{number}{3} (\bibinfo{date}{April} \bibinfo{year}{2020}),
  \bibinfo{pages}{52–55}.
\newblock
\showISSN{1072-5520}
\urldef\tempurl%
\url{https://doi.org/10.1145/3386389}
\showDOI{\tempurl}


\bibitem[\protect\citeauthoryear{Moreschi, Pereira, and Cozman}{Moreschi
  et~al\mbox{.}}{2020}]%
        {Moreschi2020}
\bibfield{author}{\bibinfo{person}{Bruno Moreschi}, \bibinfo{person}{Gabriel
  Pereira}, {and} \bibinfo{person}{Fabio~G. Cozman}.}
  \bibinfo{year}{2020}\natexlab{}.
\newblock \showarticletitle{{The Brazilian Workers in Amazon Mechanical Turk:
  Dreams and realities of ghost workers}}.
\newblock \bibinfo{journal}{\emph{Revista Contracampo}} \bibinfo{volume}{39},
  \bibinfo{number}{1} (\bibinfo{date}{apr} \bibinfo{year}{2020}).
\newblock
\showISSN{2238-2577}
\urldef\tempurl%
\url{https://doi.org/10.22409/contracampo.v39i1.38252}
\showDOI{\tempurl}


\bibitem[\protect\citeauthoryear{Muller, Lange, Wang, Piorkowski, Tsay, Liao,
  Dugan, and Erickson}{Muller et~al\mbox{.}}{2019}]%
        {muller2019}
\bibfield{author}{\bibinfo{person}{Michael Muller}, \bibinfo{person}{Ingrid
  Lange}, \bibinfo{person}{Dakuo Wang}, \bibinfo{person}{David Piorkowski},
  \bibinfo{person}{Jason Tsay}, \bibinfo{person}{Q.~Vera Liao},
  \bibinfo{person}{Casey Dugan}, {and} \bibinfo{person}{Thomas Erickson}.}
  \bibinfo{year}{2019}\natexlab{}.
\newblock \showarticletitle{How {Data} {Science} {Workers} {Work} with {Data}:
  {Discovery}, {Capture}, {Curation}, {Design}, {Creation}}. In
  \bibinfo{booktitle}{\emph{Proceedings of the 2019 {CHI} {Conference} on
  {Human} {Factors} in {Computing} {Systems}}} \emph{(\bibinfo{series}{{CHI}
  ’19})}. \bibinfo{publisher}{Association for Computing Machinery},
  \bibinfo{address}{Glasgow, Scotland Uk}, \bibinfo{pages}{1--15}.
\newblock
\showISBNx{978-1-4503-5970-2}
\urldef\tempurl%
\url{https://doi.org/10.1145/3290605.3300356}
\showDOI{\tempurl}


\bibitem[\protect\citeauthoryear{Muller, Wolf, Andres, Ashktorab, Joshi,
  Desmond, Sharma, Brimijoin, Pan, Duesterwald, and Dugan}{Muller
  et~al\mbox{.}}{2021}]%
        {muller2021}
\bibfield{author}{\bibinfo{person}{Michael Muller},
  \bibinfo{person}{Christine~T Wolf}, \bibinfo{person}{Josh Andres},
  \bibinfo{person}{Zahra Ashktorab}, \bibinfo{person}{Narendra~Nath Joshi},
  \bibinfo{person}{Michael Desmond}, \bibinfo{person}{Aabhas Sharma},
  \bibinfo{person}{Kristina Brimijoin}, \bibinfo{person}{Qian Pan},
  \bibinfo{person}{Evelyn Duesterwald}, {and} \bibinfo{person}{Casey Dugan}.}
  \bibinfo{year}{2021}\natexlab{}.
\newblock \showarticletitle{Designing {Ground} {Truth} and the {Social} {Life}
  of {Labels}}.
\newblock  (\bibinfo{year}{2021}), \bibinfo{pages}{17}.
\newblock


\bibitem[\protect\citeauthoryear{Newlands}{Newlands}{2021}]%
        {Newlands2021}
\bibfield{author}{\bibinfo{person}{Gemma Newlands}.}
  \bibinfo{year}{2021}\natexlab{}.
\newblock \showarticletitle{Lifting the curtain: {Strategic} visibility of
  human labour in {AI}-as-a-{Service}}.
\newblock \bibinfo{journal}{\emph{Big Data \& Society}} \bibinfo{volume}{8},
  \bibinfo{number}{1} (\bibinfo{date}{Jan.} \bibinfo{year}{2021}),
  \bibinfo{pages}{205395172110160}.
\newblock
\showISSN{2053-9517, 2053-9517}
\urldef\tempurl%
\url{https://doi.org/10.1177/20539517211016026}
\showDOI{\tempurl}


\bibitem[\protect\citeauthoryear{Nowicka-Franczak}{Nowicka-Franczak}{2021}]%
        {nowicka-franczak2021}
\bibfield{author}{\bibinfo{person}{Magdalena Nowicka-Franczak}.}
  \bibinfo{year}{2021}\natexlab{}.
\newblock \showarticletitle{Post-{Foucauldian} {Discourse} and {Dispositif}
  {Analysis} in the {Post}-{Socialist} {Field} of {Research}: {Methodological}
  {Remarks}}.
\newblock \bibinfo{journal}{\emph{Qualitative Sociology Review}}
  \bibinfo{volume}{17}, \bibinfo{number}{1} (\bibinfo{date}{Feb.}
  \bibinfo{year}{2021}), \bibinfo{pages}{72--95}.
\newblock
\showISSN{1733-8077}
\urldef\tempurl%
\url{https://doi.org/10.18778/1733-8077.17.1.6}
\showDOI{\tempurl}


\bibitem[\protect\citeauthoryear{Passi and Jackson}{Passi and Jackson}{2017}]%
        {passi2017}
\bibfield{author}{\bibinfo{person}{Samir Passi} {and} \bibinfo{person}{Steven
  Jackson}.} \bibinfo{year}{2017}\natexlab{}.
\newblock \showarticletitle{Data {Vision}: {Learning} to {See} {Through}
  {Algorithmic} {Abstraction}}. In \bibinfo{booktitle}{\emph{Proceedings of the
  2017 {ACM} {Conference} on {Computer} {Supported} {Cooperative} {Work} and
  {Social} {Computing}}} \emph{(\bibinfo{series}{{CSCW} ’17})}.
  \bibinfo{publisher}{Association for Computing Machinery},
  \bibinfo{address}{Portland, Oregon, USA}, \bibinfo{pages}{2436--2447}.
\newblock
\showISBNx{978-1-4503-4335-0}
\urldef\tempurl%
\url{https://doi.org/10.1145/2998181.2998331}
\showDOI{\tempurl}


\bibitem[\protect\citeauthoryear{Passi and Jackson}{Passi and Jackson}{2018}]%
        {passi2018b}
\bibfield{author}{\bibinfo{person}{Samir Passi} {and}
  \bibinfo{person}{Steven~J. Jackson}.} \bibinfo{year}{2018}\natexlab{}.
\newblock \showarticletitle{Trust in {Data} {Science}: {Collaboration},
  {Translation}, and {Accountability} in {Corporate} {Data} {Science}
  {Projects}}.
\newblock \bibinfo{journal}{\emph{Proc. ACM Hum.-Comput. Interact.}}
  \bibinfo{volume}{2}, \bibinfo{number}{CSCW} (\bibinfo{date}{Nov.}
  \bibinfo{year}{2018}), \bibinfo{pages}{1--28}.
\newblock
\showISSN{25730142}
\urldef\tempurl%
\url{https://doi.org/10.1145/3274405}
\showDOI{\tempurl}


\bibitem[\protect\citeauthoryear{Poell, Nieborg, and van Dijck}{Poell
  et~al\mbox{.}}{2019}]%
        {Poell2019}
\bibfield{author}{\bibinfo{person}{Thomas Poell}, \bibinfo{person}{David~B.
  Nieborg}, {and} \bibinfo{person}{Jos{\'{e}} van Dijck}.}
  \bibinfo{year}{2019}\natexlab{}.
\newblock \showarticletitle{{Platformisation}}.
\newblock \bibinfo{journal}{\emph{Internet Policy Review}} \bibinfo{volume}{8},
  \bibinfo{number}{4} (\bibinfo{year}{2019}).
\newblock
\urldef\tempurl%
\url{https://doi.org/10.14763/2019.4.1425}
\showDOI{\tempurl}


\bibitem[\protect\citeauthoryear{Posada}{Posada}{2020}]%
        {Posada2020a}
\bibfield{author}{\bibinfo{person}{Julian Posada}.}
  \bibinfo{year}{2020}\natexlab{}.
\newblock \showarticletitle{{The Future of Work Is Here: Toward a Comprehensive
  Approach to Artificial Intelligence and Labour}}.
\newblock \bibinfo{journal}{\emph{Ethics in Context}} (\bibinfo{year}{2020}).
\newblock
\showISSN{10185909}


\bibitem[\protect\citeauthoryear{Posada}{Posada}{2022}]%
        {Posada2022}
\bibfield{author}{\bibinfo{person}{Julian Posada}.}
  \bibinfo{year}{2022}\natexlab{}.
\newblock \showarticletitle{{Embedded Reproduction in Platform Data Work}}.
\newblock \bibinfo{journal}{\emph{Information, Communication {\&} Society}}
  (\bibinfo{year}{2022}).
\newblock


\bibitem[\protect\citeauthoryear{Posada, Weller, and Wong}{Posada
  et~al\mbox{.}}{2021}]%
        {Posada2021b}
\bibfield{author}{\bibinfo{person}{Julian Posada}, \bibinfo{person}{Nicholas
  Weller}, {and} \bibinfo{person}{Wendy~H. Wong}.}
  \bibinfo{year}{2021}\natexlab{}.
\newblock \showarticletitle{{We Haven't Gone Paperless Yet: Why the Printing
  Press Can Help Us Understand Data and AI}}.
\newblock \bibinfo{journal}{\emph{Proceedings of the 2021 AAAI/ACM Conference
  on AI, Ethics, and Society (AIES '21)}} (\bibinfo{year}{2021}).
\newblock


\bibitem[\protect\citeauthoryear{Qadri}{Qadri}{2020}]%
        {Qadri2020}
\bibfield{author}{\bibinfo{person}{Rida Qadri}.}
  \bibinfo{year}{2020}\natexlab{}.
\newblock \showarticletitle{{Algorithmized but not Atomized? How Digital
  Platforms Engender New Forms of Worker Solidarity in Jakarta}}. In
  \bibinfo{booktitle}{\emph{Proceedings of the AAAI/ACM Conference on AI,
  Ethics, and Society}}. \bibinfo{publisher}{ACM}, \bibinfo{address}{New York,
  NY, USA}, \bibinfo{pages}{144--144}.
\newblock
\showISBNx{9781450371100}
\urldef\tempurl%
\url{https://doi.org/10.1145/3375627.3375816}
\showDOI{\tempurl}


\bibitem[\protect\citeauthoryear{Raffnsøe, Gudmand-Høyer, and
  Thaning}{Raffnsøe et~al\mbox{.}}{2016}]%
        {raffnsoe2016}
\bibfield{author}{\bibinfo{person}{Sverre Raffnsøe}, \bibinfo{person}{Marius
  Gudmand-Høyer}, {and} \bibinfo{person}{Morten~S. Thaning}.}
  \bibinfo{year}{2016}\natexlab{}.
\newblock \showarticletitle{Foucault’s dispositive: {The} perspicacity of
  dispositive analytics in organizational research}.
\newblock \bibinfo{journal}{\emph{Organization}} \bibinfo{volume}{23},
  \bibinfo{number}{2} (\bibinfo{date}{March} \bibinfo{year}{2016}),
  \bibinfo{pages}{272--298}.
\newblock
\showISSN{1350-5084}
\urldef\tempurl%
\url{https://doi.org/10.1177/1350508414549885}
\showDOI{\tempurl}
\newblock
\shownote{Publisher: SAGE Publications Ltd.}


\bibitem[\protect\citeauthoryear{Raval}{Raval}{2021}]%
        {raval2021}
\bibfield{author}{\bibinfo{person}{Noopur Raval}.}
  \bibinfo{year}{2021}\natexlab{}.
\newblock \showarticletitle{Interrupting invisibility in a global world}.
\newblock \bibinfo{journal}{\emph{Interactions}} \bibinfo{volume}{28},
  \bibinfo{number}{4} (\bibinfo{date}{July} \bibinfo{year}{2021}),
  \bibinfo{pages}{27--31}.
\newblock
\showISSN{1072-5520, 1558-3449}
\urldef\tempurl%
\url{https://doi.org/10.1145/3469257}
\showDOI{\tempurl}


\bibitem[\protect\citeauthoryear{Roberts}{Roberts}{2019}]%
        {Roberts2019}
\bibfield{author}{\bibinfo{person}{Sarah~T. Roberts}.}
  \bibinfo{year}{2019}\natexlab{}.
\newblock \bibinfo{booktitle}{\emph{{Behind the Screen: Content Moderation in
  the Shadows of Social Media}}}.
\newblock \bibinfo{publisher}{Yale University Press}, \bibinfo{address}{New
  Haven, CT}. 280 pages.
\newblock
\showISSN{1461-4448}
\urldef\tempurl%
\url{https://doi.org/10.1177/1461444819878844}
\showDOI{\tempurl}


\bibitem[\protect\citeauthoryear{Ross, Irani, Silberman, Zaldivar, and
  Tomlinson}{Ross et~al\mbox{.}}{2010}]%
        {ross2010}
\bibfield{author}{\bibinfo{person}{Joel Ross}, \bibinfo{person}{Lilly Irani},
  \bibinfo{person}{M.~Six Silberman}, \bibinfo{person}{Andrew Zaldivar}, {and}
  \bibinfo{person}{Bill Tomlinson}.} \bibinfo{year}{2010}\natexlab{}.
\newblock \showarticletitle{Who are the crowdworkers?: shifting demographics in
  mechanical turk}. In \bibinfo{booktitle}{\emph{{CHI} '10 {Extended}
  {Abstracts} on {Human} {Factors} in {Computing} {Systems}}}.
  \bibinfo{publisher}{ACM}, \bibinfo{address}{Atlanta Georgia USA},
  \bibinfo{pages}{2863--2872}.
\newblock
\showISBNx{978-1-60558-930-5}
\urldef\tempurl%
\url{https://doi.org/10.1145/1753846.1753873}
\showDOI{\tempurl}


\bibitem[\protect\citeauthoryear{Salehi, Irani, Bernstein, Alkhatib, Ogbe,
  Milland, and {Clickhappier}}{Salehi et~al\mbox{.}}{2015}]%
        {salehi2015a}
\bibfield{author}{\bibinfo{person}{Niloufar Salehi}, \bibinfo{person}{Lilly~C.
  Irani}, \bibinfo{person}{Michael~S. Bernstein}, \bibinfo{person}{Ali
  Alkhatib}, \bibinfo{person}{Eva Ogbe}, \bibinfo{person}{Kristy Milland},
  {and} \bibinfo{person}{{Clickhappier}}.} \bibinfo{year}{2015}\natexlab{}.
\newblock \showarticletitle{We {Are} {Dynamo}: {Overcoming} {Stalling} and
  {Friction} in {Collective} {Action} for {Crowd} {Workers}}. In
  \bibinfo{booktitle}{\emph{Proceedings of the 33rd {Annual} {ACM} {Conference}
  on {Human} {Factors} in {Computing} {Systems}}}. \bibinfo{publisher}{ACM},
  \bibinfo{address}{Seoul Republic of Korea}, \bibinfo{pages}{1621--1630}.
\newblock
\showISBNx{978-1-4503-3145-6}
\urldef\tempurl%
\url{https://doi.org/10.1145/2702123.2702508}
\showDOI{\tempurl}


\bibitem[\protect\citeauthoryear{Sambasivan}{Sambasivan}{2022}]%
        {Sambasivan2022}
\bibfield{author}{\bibinfo{person}{Nithya Sambasivan}.}
  \bibinfo{year}{2022}\natexlab{}.
\newblock \showarticletitle{{All Equation, No Human: The Myopia of AI Models}}.
\newblock \bibinfo{journal}{\emph{Interactions}} \bibinfo{volume}{29},
  \bibinfo{number}{2} (\bibinfo{date}{mar} \bibinfo{year}{2022}),
  \bibinfo{pages}{78--80}.
\newblock
\showISSN{1072-5520}
\urldef\tempurl%
\url{https://doi.org/10.1145/3516515}
\showDOI{\tempurl}


\bibitem[\protect\citeauthoryear{Sambasivan, Kapania, Highfill, Akrong,
  Paritosh, and Aroyo}{Sambasivan et~al\mbox{.}}{2021}]%
        {Sambasivan2021}
\bibfield{author}{\bibinfo{person}{Nithya Sambasivan}, \bibinfo{person}{Shivani
  Kapania}, \bibinfo{person}{Hannah Highfill}, \bibinfo{person}{Diana Akrong},
  \bibinfo{person}{Praveen Paritosh}, {and} \bibinfo{person}{Lora~M Aroyo}.}
  \bibinfo{year}{2021}\natexlab{}.
\newblock \showarticletitle{{“Everyone wants to do the model work, not the
  data work”: Data Cascades in High-Stakes AI}}. In
  \bibinfo{booktitle}{\emph{Proceedings of the 2021 CHI Conference on Human
  Factors in Computing Systems}}. \bibinfo{publisher}{ACM},
  \bibinfo{address}{New York, NY, USA}, \bibinfo{pages}{1--15}.
\newblock
\showISBNx{9781450380966}
\urldef\tempurl%
\url{https://doi.org/10.1145/3411764.3445518}
\showDOI{\tempurl}


\bibitem[\protect\citeauthoryear{Scheuerman, Denton, and Hanna}{Scheuerman
  et~al\mbox{.}}{2021}]%
        {scheuerman2021}
\bibfield{author}{\bibinfo{person}{Morgan~Klaus Scheuerman},
  \bibinfo{person}{Emily Denton}, {and} \bibinfo{person}{Alex Hanna}.}
  \bibinfo{year}{2021}\natexlab{}.
\newblock \showarticletitle{Do {Datasets} {Have} {Politics}? {Disciplinary}
  {Values} in {Computer} {Vision} {Dataset} {Development}}.
\newblock \bibinfo{journal}{\emph{arXiv:2108.04308 [cs]}} (\bibinfo{date}{Aug.}
  \bibinfo{year}{2021}).
\newblock
\urldef\tempurl%
\url{https://doi.org/10.1145/3476058}
\showDOI{\tempurl}
\newblock
\shownote{arXiv: 2108.04308.}


\bibitem[\protect\citeauthoryear{Schteingart, Trombetta, and
  Pascuariello}{Schteingart et~al\mbox{.}}{2020}]%
        {schteingart2020}
\bibfield{author}{\bibinfo{person}{Daniel Schteingart}, \bibinfo{person}{Martin
  Trombetta}, {and} \bibinfo{person}{Gisella Pascuariello}.}
  \bibinfo{year}{2020}\natexlab{}.
\newblock \showarticletitle{Primas salariales sectoriales en {Argentina}}.
\newblock \bibinfo{journal}{\emph{Ministerio de Desarrollo Productivo de la
  Nación.}}  \bibinfo{volume}{Centro de Estudios para la Producción XXI}
  (\bibinfo{date}{Nov.} \bibinfo{year}{2020}), \bibinfo{pages}{39}.
\newblock


\bibitem[\protect\citeauthoryear{Seidelin, Dittrich, and Grönvall}{Seidelin
  et~al\mbox{.}}{2018}]%
        {seidelin2018}
\bibfield{author}{\bibinfo{person}{Cathrine Seidelin}, \bibinfo{person}{Yvonne
  Dittrich}, {and} \bibinfo{person}{Erik Grönvall}.}
  \bibinfo{year}{2018}\natexlab{}.
\newblock \showarticletitle{Data {Work} in a {Knowledge}-{Broker}
  {Organisation}: {How} {Cross}-{Organisational} {Data} {Maintenance} {Shapes}
  {Human} {Data} {Interactions}}.
\newblock \bibinfo{journal}{\emph{In Proceedings of the 32nd International BCS
  Human Computer Interaction Conference (Belfast, United Kingdom) (HCI ’18).
  BCS Learning Development Ltd., Swindon, GBR, Article 14, 12 pages. https:
  //doi.org/10.14236/ewic/HCI2018.14}} (\bibinfo{year}{2018}).
\newblock


\bibitem[\protect\citeauthoryear{Spiel}{Spiel}{2017}]%
        {spiel2017}
\bibfield{author}{\bibinfo{person}{Katta Spiel}.}
  \bibinfo{year}{2017}\natexlab{}.
\newblock \showarticletitle{Critical {Experience}: {Evaluating} (with)
  {Autistic} {Children} and {Technologies}}. In
  \bibinfo{booktitle}{\emph{Proceedings of the 2017 {CHI} {Conference}
  {Extended} {Abstracts} on {Human} {Factors} in {Computing} {Systems}}}.
  \bibinfo{publisher}{ACM}, \bibinfo{address}{Denver Colorado USA},
  \bibinfo{pages}{326--329}.
\newblock
\showISBNx{978-1-4503-4656-6}
\urldef\tempurl%
\url{https://doi.org/10.1145/3027063.3027118}
\showDOI{\tempurl}


\bibitem[\protect\citeauthoryear{Thakkar, Ismail, Kumar, Hanna, Sambasivan, and
  Kumar}{Thakkar et~al\mbox{.}}{2022}]%
        {Thakkar2022}
\bibfield{author}{\bibinfo{person}{Divy Thakkar}, \bibinfo{person}{Azra
  Ismail}, \bibinfo{person}{Pratyush Kumar}, \bibinfo{person}{Alex Hanna},
  \bibinfo{person}{Nithya Sambasivan}, {and} \bibinfo{person}{Neha Kumar}.}
  \bibinfo{year}{2022}\natexlab{}.
\newblock \showarticletitle{{When is Machine Learning Data Good ?: Valuing in
  Public Health Datafication}}.
\newblock \bibinfo{journal}{\emph{Proceedings of the 2022 CHI Conference on
  Human Factors in Computing Systems (CHI '22)}} (\bibinfo{year}{2022}).
\newblock


\bibitem[\protect\citeauthoryear{Thakkar, Kumar, and Sambasivan}{Thakkar
  et~al\mbox{.}}{2020}]%
        {Thakkar2020}
\bibfield{author}{\bibinfo{person}{Divy Thakkar}, \bibinfo{person}{Neha Kumar},
  {and} \bibinfo{person}{Nithya Sambasivan}.} \bibinfo{year}{2020}\natexlab{}.
\newblock \showarticletitle{{Towards an AI-powered Future that Works for
  Vocational Workers}}. In \bibinfo{booktitle}{\emph{Proceedings of the 2020
  CHI Conference on Human Factors in Computing Systems}}.
  \bibinfo{publisher}{ACM}, \bibinfo{address}{New York, NY, USA},
  \bibinfo{pages}{1--13}.
\newblock
\showISBNx{9781450367080}
\urldef\tempurl%
\url{https://doi.org/10.1145/3313831.3376674}
\showDOI{\tempurl}


\bibitem[\protect\citeauthoryear{Tubaro, Casilli, and Coville}{Tubaro
  et~al\mbox{.}}{2020}]%
        {Tubaro2020}
\bibfield{author}{\bibinfo{person}{Paola Tubaro}, \bibinfo{person}{Antonio~A.
  Casilli}, {and} \bibinfo{person}{Marion Coville}.}
  \bibinfo{year}{2020}\natexlab{}.
\newblock \showarticletitle{{The trainer, the verifier, the imitator: Three
  ways in which human platform workers support artificial intelligence}}.
\newblock \bibinfo{journal}{\emph{Big Data {\&} Society}} \bibinfo{volume}{7},
  \bibinfo{number}{1} (\bibinfo{year}{2020}).
\newblock
\showISSN{2053-9517}
\urldef\tempurl%
\url{https://doi.org/10.1177/2053951720919776}
\showDOI{\tempurl}


\bibitem[\protect\citeauthoryear{{United Nations}}{{United Nations}}{2015}]%
        {UnitedNations2015}
\bibfield{author}{\bibinfo{person}{{United Nations}}.}
  \bibinfo{year}{2015}\natexlab{}.
\newblock \bibinfo{title}{{Sustainable Development Goals}}.
\newblock
\newblock
\urldef\tempurl%
\url{https://www.un.org/sustainabledevelopment/}
\showURL{%
\tempurl}


\bibitem[\protect\citeauthoryear{{United Nations General Assembly}}{{United
  Nations General Assembly}}{1999}]%
        {UnitedNationsGeneralAssembly1999}
\bibfield{author}{\bibinfo{person}{{United Nations General Assembly}}.}
  \bibinfo{year}{1999}\natexlab{}.
\newblock \bibinfo{booktitle}{\emph{{Measures to combat contemporary forms of
  racism, racial discrimination, xenophobia and related intolerance}}}.
\newblock \bibinfo{type}{{T}echnical {R}eport}. \bibinfo{institution}{United
  Nations}.
\newblock


\bibitem[\protect\citeauthoryear{Wauthier and Jordan}{Wauthier and
  Jordan}{2011}]%
        {wauthier2011}
\bibfield{author}{\bibinfo{person}{Fabian~L. Wauthier} {and}
  \bibinfo{person}{Michael~I. Jordan}.} \bibinfo{year}{2011}\natexlab{}.
\newblock \showarticletitle{Bayesian {Bias} {Mitigation} for {Crowdsourcing}}.
  In \bibinfo{booktitle}{\emph{Proceedings of the 24th {International}
  {Conference} on {Neural} {Information} {Processing} {Systems}}}
  \emph{(\bibinfo{series}{{NIPS}’11})}. \bibinfo{publisher}{Curran Associates
  Inc.}, \bibinfo{address}{Granada, Spain}, \bibinfo{pages}{1800--1808}.
\newblock
\showISBNx{978-1-61839-599-3}
\urldef\tempurl%
\url{http://papers.nips.cc/paper/4311-bayesian-bias-mitigation-for-crowdsourcing.pdf}
\showURL{%
\tempurl}


\bibitem[\protect\citeauthoryear{Whelan}{Whelan}{2019}]%
        {whelan2019}
\bibfield{author}{\bibinfo{person}{Glen Whelan}.}
  \bibinfo{year}{2019}\natexlab{}.
\newblock \showarticletitle{Born {Political}: {A} {Dispositive} {Analysis} of
  {Google} and {Copyright}}.
\newblock \bibinfo{journal}{\emph{Business \& Society}} \bibinfo{volume}{58},
  \bibinfo{number}{1} (\bibinfo{date}{Jan.} \bibinfo{year}{2019}),
  \bibinfo{pages}{42--73}.
\newblock
\showISSN{0007-6503, 1552-4205}
\urldef\tempurl%
\url{https://doi.org/10.1177/0007650317717701}
\showDOI{\tempurl}


\bibitem[\protect\citeauthoryear{Wichum}{Wichum}{2013}]%
        {wichum2013}
\bibfield{author}{\bibinfo{person}{Ricky Wichum}.}
  \bibinfo{year}{2013}\natexlab{}.
\newblock \showarticletitle{Security as {Dispositif}: {Michel} {Foucault} in
  the {Field} of {Security}}.
\newblock \bibinfo{journal}{\emph{Foucault Studies}} (\bibinfo{date}{Jan.}
  \bibinfo{year}{2013}), \bibinfo{pages}{164--171}.
\newblock
\showISSN{1832-5203}
\urldef\tempurl%
\url{https://doi.org/10.22439/fs.v0i15.3996}
\showDOI{\tempurl}


\bibitem[\protect\citeauthoryear{Wood, Graham, Lehdonvirta, and Hjorth}{Wood
  et~al\mbox{.}}{2019}]%
        {Wood2019a}
\bibfield{author}{\bibinfo{person}{Alex~J. Wood}, \bibinfo{person}{Mark
  Graham}, \bibinfo{person}{Vili Lehdonvirta}, {and} \bibinfo{person}{Isis
  Hjorth}.} \bibinfo{year}{2019}\natexlab{}.
\newblock \showarticletitle{{Networked but Commodified: The (Dis)Embeddedness
  of Digital Labour in the Gig Economy}}.
\newblock \bibinfo{journal}{\emph{Sociology}} (\bibinfo{year}{2019}).
\newblock
\showISSN{00380385}
\urldef\tempurl%
\url{https://doi.org/10.1177/0038038519828906}
\showDOI{\tempurl}


\bibitem[\protect\citeauthoryear{Wood, Lehdonvirta, and Graham}{Wood
  et~al\mbox{.}}{2018}]%
        {Wood2018d}
\bibfield{author}{\bibinfo{person}{Alex~J. Wood}, \bibinfo{person}{Vili
  Lehdonvirta}, {and} \bibinfo{person}{Mark Graham}.}
  \bibinfo{year}{2018}\natexlab{}.
\newblock \showarticletitle{{Workers of the Internet unite? Online freelancer
  organisation among remote gig economy workers in six Asian and African
  countries}}.
\newblock \bibinfo{journal}{\emph{New Technology, Work and Employment}}
  \bibinfo{volume}{33}, \bibinfo{number}{2} (\bibinfo{year}{2018}),
  \bibinfo{pages}{95--112}.
\newblock
\showISSN{1468005X}
\urldef\tempurl%
\url{https://doi.org/10.1111/ntwe.12112}
\showDOI{\tempurl}


\bibitem[\protect\citeauthoryear{Woodcock and Graham}{Woodcock and
  Graham}{2020}]%
        {Woodcock2020}
\bibfield{author}{\bibinfo{person}{Jamie Woodcock} {and} \bibinfo{person}{Mark
  Graham}.} \bibinfo{year}{2020}\natexlab{}.
\newblock \bibinfo{booktitle}{\emph{{The Gig Economy: A Critical
  Introduction}}}.
\newblock \bibinfo{publisher}{Polity Press}, \bibinfo{address}{London}. 160
  pages.
\newblock


\bibitem[\protect\citeauthoryear{Zuboff}{Zuboff}{2019}]%
        {Zuboff2019}
\bibfield{author}{\bibinfo{person}{Shoshana Zuboff}.}
  \bibinfo{year}{2019}\natexlab{}.
\newblock \bibinfo{booktitle}{\emph{{The Age of Surveillance Capitalism: The
  Fight for a Human Future at the New Frontier of Power}}}.
\newblock \bibinfo{publisher}{PublicAffairs}, \bibinfo{address}{New York, NY}.
  691 pages.
\newblock


\end{thebibliography}
\end{document}